# L'APPROCHE ENERGETIQUE EN MECANIQUE DES MATERIAUX GRANULAIRES

## E. FROSSARD

Expert, Bureau d'Ingénieurs-Conseils Coyne et Bellier
frossardetienne@aol.com

### Abstract

*This communication establishes first, that basis of discontinuous granular materials mechanics, are the explicit result of a new approach of the physics of energy dissipation by friction, and second, that these mechanics of discontinua, transposed to the equivalent continuous media, as used in geomechanics, is directly responsible for a large set of practical properties of these materials, part of the fundamentals of Soil Mechanics.*

*These physics of energy dissipation are developed on the original concept of **energy rate tensor of internal actions.***

*The communication displays first:*
- *an original tensorial pattern underlying the well known laws of friction, at the elementary scale of moving contact;*
- *the macrostructural solution for the discontinuous granular mass, under general threedimensional movements, as a result of a tensorial structure associated with the distribution of moving contacts.*

*It is shown that the jump from the individual behaviour (elementary contact) to the collective behaviour (granular mass) makes appear into the general dissipation equation, a **population effect,** related to the intensity of a kind of **interaction** between moving contacts, the "internal feedback"(réalimentation interne), and related to the grade of **disorder** in the distribution of moving contacts, interaction and disorder determining the intensity of energy dissipation.*

*Among the solutions, appear those of **minimal dissipation,** associated with **ordered patterns** in the distribution of moving contacts orientations: in plane strain these ordered patterns result precisely into the RANKINE slip-lines, well known in geomechanics.*

*Concerning the equivalent continuous media, the communication draws a parallel with a dissipation equation for the continuum, established an published long ago by the author, on the basis of physical interpretation of a wide set of experimental results on these materials. This equation leads analytically to:*
- *the ROWE "stress-dilatancy" relations;*
- *the COULOMB failure criterion, at critical state;*
- *the densification by cyclical strains, in certain conditions, and other features of the "characteristic domain".*

*The structure of this dissipation equation for the equivalent continuum, is shown to be formally identical to the general dissipation equation for the discontinuous granular mass, including the presence of a particular energy rate tensor of internal actions, build on stress and strain rate tensors.This identification leads to express the macroscopic quantities, stress and strain rates,in function of microscopic variables, contact forces, sliding velocities, and geometry of packing.*

*It is shown also that experimental results suggest strongly the presence of a <u>minimum energy principle</u>: **according to limit conditions allowance, the movements into the material tend towards one of the minimal dissipation solutions**.The physical meaning of this principle, is then related to a key result of the thermodynamicians of irreversible processes (PRIGOGINE): **the theorem of minimum entropy production.***

______________________________________________________________________________________

## NOTATIONS

### LE DISCONTINU
*Contact élémentaire*

$\vec{f}(a/b)$      force de contact exercée par le grain $(a)$ sur le grain $(b)$

$\vec{v}(a/b)$      vitesse de glissement au contact, du grain $(a)$ par rapport au grain $(b)$

$\|\vec{v}\|$      norme euclidienne du vecteur $\vec{v}$

$y$      angle de friction physique moyen aux contacts entre grains

$\underline{\mathbf{p}}(a/b)$      tenseur de puissance des actions de contact élémentaire $(a/b)$ :

$$\underline{\mathbf{p}}(a/b) = \tfrac{1}{2}\left\{\vec{f}(a/b) \otimes \vec{v}(a/b) + \vec{v}(a/b) \otimes \vec{f}(a/b)\right\}$$

$\mathbf{p_i}$      valeur propre du tenseur $\underline{\mathbf{p}}$

$\mathbf{p}^+(a/b)$      "puissance reçue" au contact élémentaire:

            somme des valeurs propres *positives* du tenseur $\underline{\mathbf{p}}(a/b)$

$\mathbf{p}^-(a/b)$      "puissance rendue" au contact élémentaire:

            somme des valeurs propres *négatives* du tenseur $\underline{\mathbf{p}}(a/b)$

$Tr\{\underline{\mathbf{p}}\}$      trace du tenseur $\underline{\mathbf{p}}$

$N_\|\{\underline{\mathbf{p}}\}$      norme tensorielle du tenseur $\underline{\mathbf{p}}$, égale à la somme des valeurs absolues des valeurs propres de $\underline{\mathbf{p}}$

*Amas granulaire*

$\underline{\mathbf{P}}(A)$      tenseur de puissance des actions intérieures, de l'amas granulaire A :

            somme des tenseurs de puissance des actions de tous les contacts élémentaires contenus dans l'amas.

$\mathbf{P}^+(A)$      "puissance reçue" de l'amas granulaire:

$\mathbf{P}^-(A)$      "puissance rendue" de l'amas granulaire:

$R$      taux de réalimentation interne

$T_D$      taux de dissipation interne

### MILIEU CONTINU EQUIVALENT

$\underline{\mathbf{s}}, \underline{\dot{\mathbf{e}}}$      Tenseurs (eulériens) des contraintes et des vitesses de déformation

$\mathbb{p}$      tenseur de puissance des actions intérieures, pour le milieu continu :

$$\mathbf{p} = \tfrac{1}{2}\left\{\underline{\underline{\mathbf{s}}} \otimes \underline{\underline{\dot{\mathbf{e}}}} + \underline{\underline{\dot{\mathbf{e}}}} \otimes \underline{\underline{\mathbf{s}}}\right\} \textit{ (produit contracté)}$$

## CONVENTIONS ET HYPOTHESES DE BASE

Dans ce qui suit, nous analyserons le mouvement, dans des référentiels orthonormés, de matériaux granulaires dans les conditions suivantes, au voisinage de l'équilibre statique :

1) le matériau est soumis à des déformations lentes, résultant de mouvements localisés aux contacts entre les grains, ces contacts sont unilatéraux, et purement frictionnels; les coefficients de friction sur ces contacts sont suffisamment voisins pour être représentés par une valeur moyenne unique.

2) les grains sont indéformables, de dimensions et de formes aléatoires, mais suffisamment proches de la convexité pour que:
   -les zones de contact entre deux grains puissent être raisonnablement schématisées par un seul point de contact;
   -le matériau granulaire dans son ensemble ne présente aucune résistance à la traction localement ou globalement (condition de non-traction);

3) l'énergie dissipée dans les contacts est due aux seuls mouvements relatifs de glissement, l'effet des mouvements de pivotement et de roulement est négligeable; la représentation du torseur des efforts de contact par sa seule résultante, est suffisante;

4) les mouvements sont suffisamment lents pour que les effets dynamiques soient négligeables, ainsi que ceux de l'énergie cinétique;

5) si le matériau granulaire est baigné par un fluide remplissant les vides intergranulaires, la pression du fluide est prise comme origine des pressions (raisonnement sur les forces intergranulaires, ou les contraintes effectives), et l'on considère des mouvements suffisamment lents dans le matériau pour que les échanges d'énergie avec le fluide demeurent négligeables (la contrainte d'interface demeure la simple pression du fluide);

6) dans la représentation eulérienne du milieu continu équivalent, contraintes de compression et déformations de contraction seront notées positivement, suivant les usages en géomécanique.

Les valeurs locales des contraintes et vitesses de déformations résultent:
- d'une composante moyenne, sur laquelle s'exercent les gradients de grande échelle dus aux actions extérieures (par exemple la pesanteur);
- d'une composante de fluctuations locales, de nature aléatoire, due à l'hétérogénéité inhérente du milieu;

Ces fluctuations locales sont supposées suffisamment décorrélées pour ne pas avoir d'incidence sur la puissance des actions intérieures, et plus précisément telles que les covariances entre contraintes et vitesses de déformation soient antisymétriques.

# INTRODUCTION

Le rôle structurant de la friction physique dans le comportement mécanique à grande échelle des matériaux granulaires, modèle de référence en géomécanique, a suscité de nombreux développements depuis sa mise en évidence dans le mémoire de C.A.COULOMB(1773) [1].
Ces développements, appuyés sur les travaux fondateurs de RANKINE (1857) [2], PRANDTL (1920), CAQUOT (1934) [3], TERZAGHI (1943) [4] et bien d'autres encore, se sont peu à peu intégrés dans le corps de connaissances de la Mécanique des Sols, un des piliers des sciences de l'ingénieur dans le domaine du Génie Civil. Ces développements continuent dans certains secteurs de la recherche fondamentale actuelle [5][6], qui s'attachent par ailleurs aux questions intriguantes de physique de la matière "ordinaire", ou de la matière divisée, que posent ces matériaux [7].

Toutefois le lien direct entre les causes premières –*la friction aux contacts entre les grains*- et l'ensemble des manifestations globales –*le critère de rupture et autres aspects du comportement mécanique tridimensionnel à grande échelle du matériau*- s'il a fait l'objet d'explorations, dont celle de O.REYNOLDS(1885)[8], et d'avancées remarquables comme la théorie "contraintes-dilatance" de ROWE(1962)[9], reprise et développée par HORNE (1965-1969) [10],[11], demeure sans clarification aboutie, de portée générale.

Dans ces pages, nous verrons que ce lien peut être complètement établi, à travers une physique explicite: celle de la dissipation d'énergie, dans des mouvements éventuellement importants, mais demeurant proches des conditions d'équilibre statique.
Cette physique est fondée sur un concept original que nous verrons s'imposer naturellement, celui des *"tenseurs de puissance des actions intérieures"*, déjà noté dans une publication antérieure sur ces sujets.

Pour la clarté de l'exposé, nous établirons d'abord cette physique pour le matériau discontinu, depuis le contact élémentaire jusqu'à l'amas granulaire dans son mouvement à trois dimensions, avant d'en aborder les conséquences sur les propriétés du milieu continu équivalent.

Au contact élémentaire, *Chapitre 1*, les lois bien connues de la friction révèlent une structure tensorielle simple, celle du tenseur de puissance des actions de contact, qui réunit de façon condensée la géométrie des forces et des mouvements, avec la dissipation d'énergie qu'ils engendrent.
Les propriétés remarquables de ce tenseur de puissance, et leur signification physique, structurent l'ensemble des résultats que nous établirons ensuite:
- les notions objectives de *"puissance reçue"* et *"puissance rendue"*, qui font apparaître le contact élémentaire comme une sorte de transformateur, recevant de l'énergie d'un côté, en rendant de l'autre, tout en ayant dissipé l'essentiel au passage, le tout en proportions fixées;
- la forme condensée de l'équation de dissipation, dotée de propriétés algébriques fortes, qui sera la clé de l'amas granulaire.

Dans l'amas granulaire, *Chapitre 2*, la sommation des tenseurs de puissance élémentaires, forme le tenseur de puissance global de l'amas, qui reliera les propriétés d'ensemble de la population des contacts en mouvement, à celle de ses "individus" élémentaires.
Cette description tensorielle simple, qui mène à établir l'équation de dissipation pour l'amas granulaire, permet l'analyse du mouvement général à trois dimensions, et la mise en évidence d'un effet de population**:** la *"réalimentation interne"*, par lequel les contacts élémentaires interagissent en échangeant de l'énergie, et qui influe sur le bilan énergétique global de l'amas.
Nous verrons que cet effet de population structure l'ensemble des solutions du mouvement général de l'amas granulaire, entre deux limites**:**
- les solutions de dissipation minimale, possédant des structures ordonnées dans la population des contacts, qui empêchent cette réalimentation interne; parmi celles-ci apparaît une solution remarquable en déformation plane, dont la structure ordonnée correspond précisément aux "lignes de glissement de Rankine";
- les solutions de dissipation maximale, dans lesquelles le désordre dans la population des contacts, maximise la réalimentation interne.
Enfin nous mettrons en évidence un ensemble de propriétés essentielles des solutions de l'équation générale de la dissipation pour l'amas granulaire, et leurs correspondance avec des caractères observés dans le comportement usuel de ces matériaux.

Nous aborderons ensuite les conséquences sur les propriétés du milieu continu équivalent, *Chapitre 3*, à partir de résultats antérieurs de l'auteur, établis par l'interprétation physique du comportement mécanique expérimental de ces matériaux [12] [13]: une équation de dissipation du continu équivalent, qui contient les relations de ROWE, le critère de COULOMB, l'effet de densification par mouvements alternés et autres propriétés du "domaine caractéristique"[14].
Nous verrons que cette équation de dissipation du continu, d'origine expérimentale, qui lie les valeurs propres d'un certain tenseur de puissance des actions intérieures du continu équivalent, présente une similitude complète avec la théorie du discontinu qui précède, validant l'identification énergétique entre l'amas granulaire discontinu et sa représentation continue équivalente. Cette identification énergétique nous conduira à l'expression explicite des grandeurs macroscopiques, les tenseurs de contraintes, de puissance, et des vitesses de déformation, en fonction des variables microscopiques, forces de contact, vitesses de glissement, et paramètres géométriques de l'empilement granulaire.

Nous constaterons que les résultats expérimentaux, par une certaine indifférence aux conditions initiales, dénotent des mouvements proches de la dissipation minimale, suggérant la présence d'un *principe de moindre énergie: lorsque les conditions aux limites le permettent, le matériau tend vers la dissipation minimale.*

Au *Chapitre 4*, après avoir montré, que l'idée de ce principe de moindre énergie était déjà en filigrane, voire clairement débattue, dans certains travaux fondateurs, nous établirons sa correspondance avec un résultat général établi par les thermodynamiciens de l'irréversible dont I.PRIGOGINE (1977)[15][16][17]: *le théorème de la production minimale d'entropie*, confirmant ainsi le lien entre la présente approche énergétique d'une part, et d'autre part la physique statistique et la thermodynamique de l'irréversible.

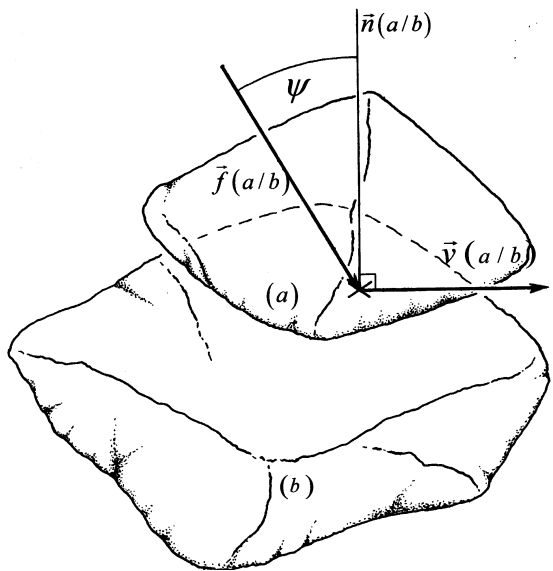

*Figure 1-1 :* Contact élémentaire

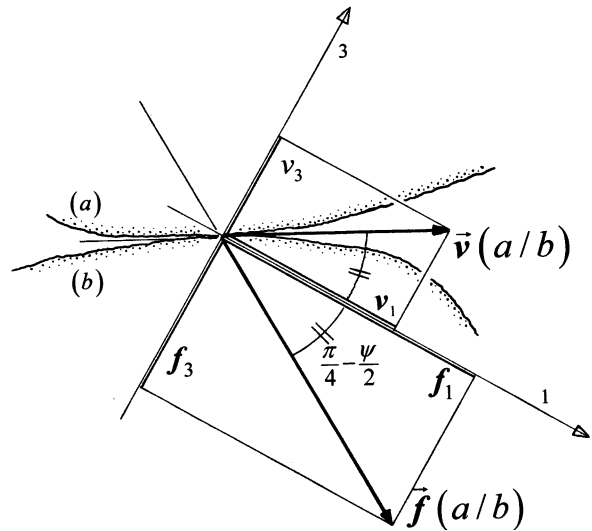

*Figure 1-2 :* Tenseur de puissance des actions de contact

## 1. LE CONTACT ELEMENTAIRE

### 1.1 ASPECTS VECTORIELS DE LA DISSIPATION D'ENERGIE

Au contact entre deux grains $(a)$ et $(b)$, Figure 1-1, l'angle entre la force de contact $\vec{f}(a/b)$, et la normale au plan de contact, est limité par l'angle de friction $\psi$, qui limite les conditions d'équilibre.

Lorsque, tout en demeurant près des conditions d'équilibre, le contact se met en mouvement, la vitesse relative de glissement (*notons qu'il s'agit bien d'une grandeur objective*) $\vec{v}(a/b)$ prend la direction de la composante tangentielle de $\vec{f}(a/b)$, et la puissance développée par la force de contact, entièrement dissipée par friction, est donnée par le produit scalaire $\vec{f}(a/b)\cdot\vec{v}(a/b)$.

La disposition géométrique de $\vec{f}(a/b)$ et $\vec{v}(a/b)$ vérifie alors:

$$\vec{f}(a/b)\cdot\vec{v}(a/b) = \sin\gamma \cdot \|\vec{f}(a/b)\| \cdot \|\vec{v}(a/b)\| \tag{1.1}$$

Cette équation, qui demeure vérifiée lorsque le contact est au repos (la vitesse de glissement est alors nulle), exprime l'égalité entre la puissance développée par la force de contact -*au premier membre*-, et une fonction toujours positive, qui est donc une fonction de dissipation, -*au second membre*-. L'équation (1.1) est donc la formulation vectorielle de la dissipation d'énergie par friction, pour le contact élémentaire.

L'équation vectorielle (1.1) présente aussi de remarquables symétries, qui suggèrent une structure tensorielle sous-jacente.

## *1-2 ASPECTS TENSORIELS DE LA DISSIPATION D'ENERGIE: TENSEUR DE PUISSANCE*

### *1.2.1 Formulation tensorielle de la dissipation*

Considérons le tenseur symétrique, que l'on appellera dans ce qui suit "tenseur de puissance des actions de contact élémentaire":
$\underline{\mathbf{p}}(a/b) = \frac{1}{2}\{\vec{f}(a/b)\otimes\vec{v}(a/b) + \vec{v}(a/b)\otimes\vec{f}(a/b)\}$, de composantes $p_{i,j} = \frac{1}{2}\{f_i\cdot v_j + v_i\cdot f_j\}$; qui vérifie $Tr\{\underline{\mathbf{p}}(a/b)\} = \vec{f}(a/b)\cdot\vec{v}(a/b)$.

Ce tenseur est aussi symétrique par rapport à ses arguments (car $\underline{\mathbf{p}}(a/b) = \underline{\mathbf{p}}(b/a)$), ce qui montre qu'il s'agit d'une grandeur physique attachée au contact entre les deux grains, indépendamment de l'ordre dans lequel ceux-ci sont pris en compte ( *a/b*, ou *b/a*). Ce tenseur est diagonalisé dans le repère propre suivant, figure 1-2, : les axes 1 et 3 sont formés par les deux bissectrices des directions des vecteurs $\vec{f}(a/b)$ et $\vec{v}(a/b)$, l'axe 2 complète, formant un repère direct.

Dans ce repère, le tenseur $\underline{\mathbf{p}}(a/b)$ est donné par:

$$\underline{\mathbf{p}}(a/b) = \|\vec{f}(a/b)\| \cdot \|\vec{v}(a/b)\| \cdot \begin{bmatrix} \cos^2\left(\frac{\pi}{4} - \frac{\gamma}{2}\right) & 0 & 0 \\ 0 & 0 & 0 \\ 0 & 0 & -\sin^2\left(\frac{\pi}{4} - \frac{\gamma}{2}\right) \end{bmatrix} \tag{1.2}$$

Notons que les valeurs propres vérifient :

$$\begin{cases} \mathbf{p_1} + \mathbf{p_2} + \mathbf{p_3} = \vec{f} \cdot \vec{v} \\ \mathbf{p_2} = 0 \\ |\mathbf{p_1}| + |\mathbf{p_2}| + |\mathbf{p_3}| = \|\vec{f}\| \cdot \|\vec{v}\| \end{cases} \quad (1.3)$$

Rapprochant (1.3) de (1.1), on constate que la formulation vectorielle de la dissipation d'énergie, donnée précédemment, se traduit par une formulation tensorielle, où ne figure plus que le seul tenseur de puissance, par ses valeurs propres:

$$\begin{cases} \mathbf{p_1} + \mathbf{p_2} + \mathbf{p_3} = \sin y \cdot \{|\mathbf{p_1}| + |\mathbf{p_2}| + |\mathbf{p_3}|\} \\ \mathbf{p_2} = 0 \qquad \text{(ou encore } \mathbf{p_1}\mathbf{p_2}\mathbf{p_3} = 0\text{)} \end{cases} \quad (1.4)$$

Cette formulation est également vérifiée lorsque le contact est au repos, la vitesse de glissement et donc le tenseur $\underline{\mathbf{p}}(a/b)$ étant alors nuls.

## *1.3 SIGNIFICATION PHYSIQUE*

Si l'on considère le contact élémentaire comme un "système physique" échangeant de l'énergie avec l'extérieur, diverses grandeurs liées au tenseur de puissance, dont les composantes peuvent être interprétées comme des flux d'énergie mécanique échangés avec l'extérieur, prennent alors une signification remarquable.

### *1.3.1 Puissance reçue et puissance rendue*

Désignons par $\mathbf{p}^+$ et $\mathbf{p}^-$ les grandeurs objectives définies, dans le repère propre, par:

$$\begin{cases} \mathbf{p}^+ = \tfrac{1}{2}\left( \sum_i \mathbf{p_i} + \sum_j |\mathbf{p_j}| \right) \\ \mathbf{p}^- = \tfrac{1}{2}\left( \sum_i \mathbf{p_i} - \sum_j |\mathbf{p_j}| \right) \end{cases} \quad (1.4)$$

Nous appellerons "puissance reçue" la grandeur $\mathbf{p}^+$, par définition somme des valeurs propres *positives* du tenseur $\underline{\mathbf{p}}(a/b)$ (se réduisant ici à la seule $\mathbf{p_1}$), et "puissance rendue" la grandeur $\mathbf{p}^-$, par définition somme des valeurs propres *négatives* du tenseur $\underline{\mathbf{p}}(a/b)$ (se réduisant ici à la seule $\mathbf{p_3}$), pour les raisons suivantes, Figure 1-3.

Découpons par la pensée autour du contact en mouvement, un petit volume parallélipipédique dans la matière des grains, dont les faces sont parallèles aux plans du repère propre. Considérons ce volume comme un milieu pseudo-continu, vu de l'extérieur:
- des contraintes *moyennes* **s** s'exercent sur ses faces, en équilibre avec la force de contact $\vec{f}(a/b)$;
- il présente des vitesses *moyennes* de déformation **ė**, correspondant à la vitesse de glissement $\vec{v}(a/b)$;
- ces contraintes et vitesses de déformation moyennes peuvent s'exprimer en fonction de $\vec{f}(a/b)$ et $\vec{v}(a/b)$ (voir Annexe 1-1).

Notre volume pseudo-continu échange avec l'extérieur, les flux d'énergie apparents suivants:
- il reçoit de l'énergie mécanique suivant l'axe 1, la puissance *reçue*, produit du volume par la contrainte (principale) $\mathbf{s_1}$ et la vitesse de déformation (principale) $\mathbf{\dot{e}_1}$, est précisément égale à $\mathbf{p_1} = \mathbf{p}^+$ ;
- il ne reçoit ni rend d'énergie suivant l'axe 2;
- il rend de l'énergie mécanique suivant l'axe 3, la puissance *rendue* est précisément égale à $\mathbf{p_3} = \mathbf{p}^-$.

Avec ces définitions, la formulation tensorielle (1.4) permet d'écrire :

$$Tr\{\underline{\mathbf{p}}\} = \mathbf{p}^+ + \mathbf{p}^- = \sin\mathbf{y} \cdot \{\mathbf{p}^+ - \mathbf{p}^-\} \quad (1.5)$$

, ou encore :

$$\frac{\mathbf{p}^+}{\mathbf{p}^-} = -\frac{1 + \sin\mathbf{y}}{1 - \sin\mathbf{y}} = C^{te} \quad (1.6)$$

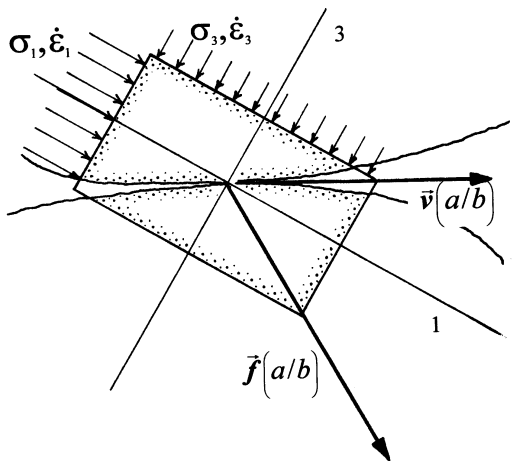

***Figure 1-3:*** *Volume pseudo-continu équivalent*

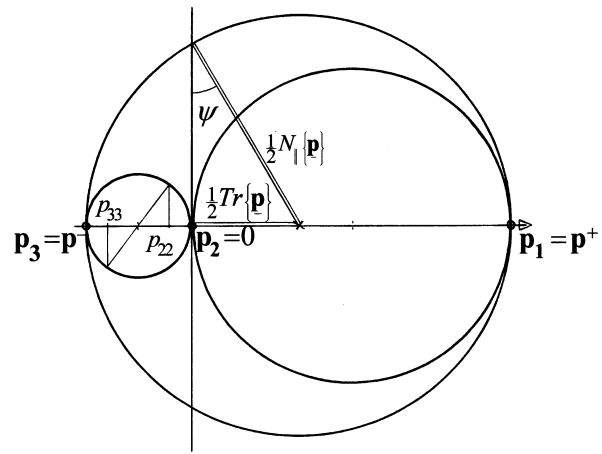

***Figure 1-4:*** *Représentation du tenseur de puissance dans le plan de Mohr*

**Au cours du mouvement, la puissance développée par les efforts de contact (égale à $Tr\{\underline{\mathbf{p}}\} = \mathbf{p}^+ + \mathbf{p}^-$), peut être décomposée en deux composantes objectives: "puissance reçue" $\mathbf{p}^+$ et "puissance rendue" $\mathbf{p}^-$, qui sont en _rapport constant_. La différence entre les valeurs absolues de ces deux composantes (c'est à dire leur somme algébrique), donne la puissance dissipée au contact.**

Le tenseur $\underline{\mathbf{p}}$, qui est un tenseur plan, peut être utilement figuré dans la représentation des cercles de Mohr, Figure 1-4.

### 1.3.2 La Norme des flux

La première des relations (1.4) relie :
- au premier membre, la trace du tenseur $\underline{\mathbf{p}}$, égale à la puissance développée par les efforts de contact, qui est aussi le"*Bilan d'entrée/sortie*" sur les échanges de notre "système physique"avec l'extérieur;
- au second membre, multipliée par le coefficient matériel $\sin y$, une fonction symétrique, toujours positive, des valeurs propres du tenseur $\underline{\mathbf{p}}$, qui possède les propriétés algébriques d'une _norme_ pour les tenseurs symétriques tels que $\underline{\mathbf{p}}(a/b)$, et dont les propriétés élémentaires, qui structurent l'ensemble de cette approche énergétique, sont présentées en Annexe 1-2; nous noterons cette fonction $N_\parallel\{\underline{\mathbf{p}}\}$;
- cette fonction peut être interprétée comme une mesure de l'intensité globale des échanges d'énergie avec l'extérieur, indépendamment de leur signe, que l'on peut désigner par "*Norme des flux*".

Avec ces définitions, la première des relations (1.4), s'énonce: *le Bilan d'entrée/sortie est proportionnel à la Norme des flux.* Ces relations (1.4) prennent la forme condensée:

$$\begin{cases} Tr\{\underline{\mathbf{p}}\} = \sin y \cdot N_\parallel\{\underline{\mathbf{p}}\} \\ D\acute{e}t\{\underline{\mathbf{p}}\} = 0 \end{cases} \quad (1.7)$$

Les relations (1.5), deviennent:

$$\begin{cases} \mathbf{p}^+ = \tfrac{1}{2}\left[Tr\{\underline{\mathbf{p}}\} + N_\parallel\{\underline{\mathbf{p}}\}\right] \\ \mathbf{p}^- = \tfrac{1}{2}\left[Tr\{\underline{\mathbf{p}}\} - N_\parallel\{\underline{\mathbf{p}}\}\right] \end{cases} \quad (1.8)$$

### 1.3.3 *Rupture de symétrie*

Les relations (1.8), jointes à la définition du tenseur $\underline{\mathbf{p}}$, renferment de manière condensée, la structure tensorielle énergétique attachée aux lois élémentaires de la friction. Les particularités de ces relations vont donc charpenter les développements des sections qui suivent.

Résultant directement de l'énergie, ces relations décrivent une symétrie apparemment parfaite entre les forces et les vitesses.

Or, les lois élémentaires de la friction comportent par ailleurs une condition qui brise cette apparente symétrie: l'unilatéralité des contacts, qui impose que la composante normale de la force de contact soit toujours en compression (ou nulle).

C'est pourquoi il est nécessaire de compléter les relations (1.8) par une condition qui représente cette rupture de symétrie: c'est le role de la condition de non-traction (hypothèse de base N° 2), qui prendra une place essentielle dans le schéma du milieu continu équivalent.

## 1.4 PROPRIETES ALGEBRIQUES

### 1.4.1 *Conservation par rotation autour des axes portant $\mathbf{p}^+$ ou $\mathbf{p}^-$, axe neutre*

Effectuons un changement de repère par rotation autour de l'axe portant $\mathbf{p}^+$, et montrons que la première des relations (1.4) demeure vérifiée par les composantes <u>diagonales</u> du tenseur $\underline{\mathbf{p}}$ dans le nouveau repère. En effet, $\mathbf{p}^+ = p_{11}$ est conservée, et, suivant les propriétés du cercle de Mohr -Figure 1-4, les nouvelles composantes $p_{22}$ et $p_{33}$ vérifient:

$$\begin{cases} p_{22} + p_{33} = \mathbf{p_2} + \mathbf{p_3} = \mathbf{p}^- \\ |p_{22}| + |p_{33}| = |\mathbf{p_2}| + |\mathbf{p_3}| = -\mathbf{p}^- \end{cases}$$

De même, dans un changement de repère par rotation autour de l'axe de $\mathbf{p}^-$.

Par contre, cette propriété n'est plus vérifiée après rotation du repère autour de l'axe 2 correspondant à la valeur propre nulle du tenseur $\underline{\mathbf{p}}$. Nous désignerons dorénavant cet axe 2, qui correspond à une puissance nulle, par "axe neutre".

### 1.4.2 *Directions de puissance tensorielle purement tangentielle*

Il est utile pour la suite d'analyser pour quelles directions de l'espace, la puissance tensorielle est purement tangentielle. En utilisant l'expression du tenseur de puissance dans le repère propre donnée en (1.2), on trouve que les seules directions (normales) pour lesquelles la puissance est purement tangentielle, sont celles du plan formé par la normale au plan de contact et l'axe 2, ainsi que celles du plan symétrique du précédent par rapport aux axes 1 ou 3, comme on peut également le constater sur le diagramme de Mohr, Figure 1-4.

Autrement dit, les "facettes" correspondantes sont deux faisceaux de plans, l'un pivotant autour de la direction de la vitesse $\vec{v}(a/b)$, l'autre pivotant autour de la direction de la force de contact $\vec{f}(a/b)$.

## 2  L'AMAS GRANULAIRE

### *2.1  INTRODUCTION*

L'amas granulaire A qui se déforme sous l'effet d'efforts extérieurs, Figure 2-1, peut être considéré comme constitué de morceaux continus (les grains, et le fluide intersticiel ), séparés par des discontinuités (les contacts, et interfaces grains/fluide).

La mécanique des milieux avec surfaces de discontinuités (voir par exemple [18]), montre que la puissance des efforts intérieurs est la somme de deux termes :
- la puissance de déformation *dans* les morceaux continus;
- la puissance développée *sur* les discontinuités, par les efforts de contact.

Dans le cadre de nos hypothèses de base, tous les termes sont négligeables, sauf la puissance développée par les actions de contact entre les grains: la puissance des efforts intérieurs pour l'amas granulaire est donc la somme des puissances développées aux contacts élémentaires.

Ceci mène donc à considérer l'amas granulaire comme une population de contacts en mouvement, et permet de restreindre le champ du problème de mécanique statistique posé par l'amas granulaire.

Par souci de clarté, nous nous limiterons aux simples sommations discrètes, qui apparaissent naturellement dans le problème de l'amas granulaire, sans faire intervenir explicitement les distributions statistiques sous-jacentes.

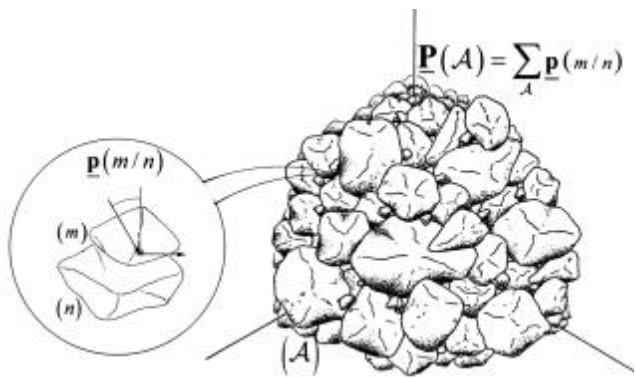

*Figure2-1  L'amas granulaire*

Cette représentation "minimaliste" ne doit cependant pas faire oublier la nature essentiellement statistique du sujet.

## *2.2 ASPECTS VECTORIELS DE LA DISSIPATION*

Numérotant de 1 à *N* les grains de l'amas A , la formulation vectorielle de la dissipation dans l'amas granulaire, s'obtient par sommation à partir de l'expression élémentaire (1.1):

$$\sum_{m<n\leq N} \vec{f}(m/n) \cdot \vec{v}(m/n) = \sin\boldsymbol{y} \cdot \sum_{m<n\leq N} \left\|\vec{f}(m/n)\right\| \cdot \left\|\vec{v}(m/n)\right\| \quad (2.1)$$

La propriété de symétrie par rapport aux arguments (m) et (n) (sections 1.1 et 1.2.1), assure que le résultat de la sommation est indépendant de l'ordre de numérotation des grains, à condition que chaque contact soit pris en compte une fois et une seule fois, c'est la raison de la sommation pour $m < n \leq N$.

Cette observation souligne que la sommation est bien attachée à la population des contacts, toutefois nous garderons la notation relative aux grains, pour la continuité de l'exposé.

## *2.3 ASPECTS TENSORIELS DE LA DISSIPATION*

### *2.3.1 Tenseur de puissance de l'amas granulaire*

L'équation de dissipation (2.1) fait intervenir des grandeurs vectorielles locales, forces et vitesses de glissement aux contacts élémentaires, dont les sommations sur l'amas granulaire n'ont pas de signification physique directe comme grandeurs globales, relatives à l'amas granulaire pris dans son ensemble, car ces grandeurs ne sont en général pas *extensives*.

Pour reformuler cette équation avec des grandeurs globales, observons que l'énergie dissipée est une grandeur extensive, et faisons apparaître les tenseurs de puissance, ainsi que leur signification physique à la fois locale et globale.

A cet effet, notons que le premier membre de (2.1) est la *trace* d'un tenseur, égal à la sommation sur l'amas granulaire, de tous les tenseurs de puissance des actions de contacts élémentaires, que nous appellerons "tenseur de puissance des actions intérieures de l'amas granulaire" $\underline{\mathbf{P}}(\mathrm{A})$ :

$$\underline{\mathbf{P}}(\mathrm{A}) = \sum_{m<n\leq N} \underline{\mathbf{p}}(m/n) \quad (2.2)$$

Le tenseur $\underline{\mathbf{P}}(\mathrm{A})$ est un tenseur symétrique du 2ème ordre, possèdant un repère propre, dans lequel on peut définir la "puissance reçue par l'amas granulaire" $\mathbf{P}^+(\mathrm{A})$ et la "puissance rendue par l'amas granulaire" $\mathbf{P}^-(A)$, à partir des valeurs propres du tenseur $\underline{\mathbf{P}}(A)$, par les expressions (1.5).

Le premier membre de (2.1) est relié à $\mathbf{P}^+(A)$ et $\mathbf{P}^-(A)$, par les définitions (1.5):

$$Tr\{\underline{\mathbf{P}}(A)\} = \mathbf{P}^+(A) + \mathbf{P}^-(A)$$

De plus, les propriétés de la norme tensorielle $N_1$ qui intervient dans ces définitions (voir section 1.3.2 et Annexe 1-2), relient $\mathbf{P}^+(A)$ et $\mathbf{P}^-(A)$, aux puissances reçues et rendues aux contacts élémentaires $\mathbf{p}^+(m/n)$ et $\mathbf{p}^-(m/n)$, par les inégalités:

$$\begin{cases} 0 \leq \mathbf{P}^+(A) \leq \sum_{m<n\leq N} \mathbf{p}^+(m/n) \\ \sum_{m<n\leq N} \mathbf{p}^-(m/n) \leq \mathbf{P}^-(A) \leq 0 \end{cases} \quad (2.3)$$

L'équation de la dissipation dans l'amas granulaire (2.1), s'écrit alors, tenant compte de (1.6):

$$Tr\{\underline{\mathbf{P}}(A)\} = \sin\boldsymbol{y} \cdot \sum_{m<n\leq N} \{\mathbf{p}^+(m/n) - \mathbf{p}^-(m/n)\} \quad (2.4)$$

Du fait des inégalités (2.3), les relations (2.2) à (2.4) n'établissent pas d'expression directe de la dissipation d'énergie dans l'amas granulaire, en fonction des seules grandeurs globales définies à partir du tenseur $\underline{\mathbf{P}}$.

Le passage du comportement individuel (le contact élémentaire) au comportement d'ensemble (l'amas granulaire), amène donc une ***indétermination physique*** (les inégalités (2.3)).

### *2.3.2 La réalimentation interne*

Montrons maintenant que cette indétermination physique est due à une certaine forme d'interaction possible entre les contacts élémentaires voisins. On montrera ensuite que l'ampleur globale de ces interactions:
- règle l'intensité de la dissipation dans l'amas granulaire, entre deux limites;
- est directement liée au degré de désordre régnant dans la distribution spatiale des orientations des contacts en mouvement.

En effet, dans l'amas granulaire en mouvement, le déplacement d'un grain a des incidences sur les mouvements de ses voisins (en bref, une pierre pousse l'autre..), et ainsi de suite en cascade. Toutefois, cet effet en chaîne peut ne pas s'étendre indéfiniment, chaque mouvement élémentaire pouvant dissiper au passage une partie de l'énergie, qu'il ne transmet donc pas au mouvement suivant, atténuant ainsi rapidement l'effet en cascade au cours de sa propagation.

Ceci peut s'exprimer au moyen des tenseurs de puissance -Figure 2-2-:

- la puissance *rendue* par un contact $\mathbf{p}^-(a/b)$ peut **réalimenter** la puissance *reçue* par un autre contact voisin $\mathbf{p}^+(b/c)$;
- ces grandeurs étant orientées, une telle réalimentation sera d'autant plus significative que la direction portant $\mathbf{p}^-(a/b)$ sera proche de la direction portant $\mathbf{p}^+(b/c)$, comme dans l'exemple de la Figure 2-2;
- dans le bilan d'ensemble des deux contacts voisins, la puissance reçue globale sera inférieure à la somme des puissances reçues élémentaires, du fait de cette réalimentation interne (et symétriquement, pour la puissance rendue); il en est de même pour le bilan d'ensemble à l'échelle de l'amas granulaire;
- les inégalités (2.3) en sont la conséquence, sur la somme des tenseurs élémentaires.

Notons que la Figure 2-2 suggère que la réalimentation interne est une interaction mutuelle: si le contact (a/b) réalimente le contact (b/c), alors le contact (b/c) réalimente aussi le contact (a/b), quoique avec une intensité différente.

On peut montrer que cette apparente symétrie dans l'interaction est effectivement vérifiée à deux dimensions, cas de la Figure 2-2, et qu'**elle ne l'est pas à trois dimensions**.

Remarquons enfin que cette notion de réalimentation s'applique aussi entre des sous-amas, formant une partition de l'amas granulaire A, car le tenseur de puissance de l'amas est toujours la somme des tenseurs de puissance des sous-amas formant la partition.

Les relations (2.2) à (2.4) permettent de définir globalement dans l'amas granulaire un "taux de réalimentation interne" $R(A)$ :

$$R(A) = \frac{\mathbf{P}^+(A) - \sum_{m<n\leq N} \mathbf{p}^+(m/n)}{\sum_{m<n\leq N} \mathbf{p}^-(m/n)} \qquad (2.5)$$

D'après les inégalités (2.3), $R(A)$ est toujours compris entre 0 et 1.

Cette première définition présente l'inconvénient de mêler des grandeurs relatives à l'amas dans son ensemble et des sommations sur les grandeurs élémentaires. Elle peut être transformée en une simple fonctionnelle statistique sur la distribution des tenseurs de puissance élémentaires, en utilisant (2.5),(2.2) et (1.8):

$$R = \frac{1}{(1-\sin y)}\left[1 - \frac{N_\parallel \left\{\sum_{m<n\leq N} \underline{\mathbf{p}}(m/n)\right\}}{\sum_{m<n\leq N} N_\parallel \left\{\underline{\mathbf{p}}(m/n)\right\}}\right] \quad (2.6)$$

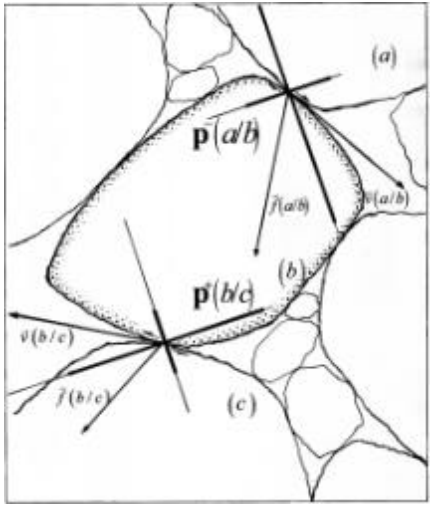

*Figure 2-2 : La réalimentation interne*
$N_\parallel \left\{\underline{\mathbf{p}}(a/b) + \underline{\mathbf{p}}(b/c)\right\} < N_\parallel \left\{\underline{\mathbf{p}}(a/b)\right\} + N_\parallel \left\{\underline{\mathbf{p}}(b/c)\right\}$

La définition (2.5), ainsi que les inégalités (2.3), conduisent alors à:

$$\begin{cases} 0 \leq R \leq 1 \\ \mathbf{P}^+(A) = \sum_{m<n\leq N} \mathbf{p}^+(m/n) + R \sum_{m<n\leq N} \mathbf{p}^-(m/n) \\ \mathbf{P}^-(A) = (1-R) \sum_{m<n\leq N} \mathbf{p}^-(m/n) \end{cases} \quad (2.7)$$

Retournant les relations (2.7) et les portant dans (2.4), l'équation de la dissipation devient:

$$Tr\{\underline{\mathbf{P}}(A)\} = \mathbf{P}^+(A) + \mathbf{P}^-(A) = \sin y \cdot \left\{\mathbf{P}^+(A) - \frac{(1+R)}{(1-R)}\mathbf{P}^-(A)\right\}, \text{ si } R<1$$

## 2.4 EQUATION GENERALE DE LA DISSIPATION

En reprenant les définitions de $\mathbf{P}^+(A)$ et $\mathbf{P}^-(A)$ en fonction de $Tr\{\underline{\mathbf{P}}(A)\}$ et $N_\parallel\{\underline{\mathbf{P}}(A)\}$, la dernière formulation ci-dessus peut être mise sous une forme plus condensée, qui mène à l'équation de la dissipation dans l'amas granulaire, dans le cas général, en fonction de seules grandeurs physiques globales $\underline{\mathbf{P}}(A)$ et $R(A)$, relatives à l'amas granulaire A dans son ensemble    (2.8):

$$\begin{cases} \textit{Equation de dissipation}: \ Tr\{\underline{\mathbf{P}}(A)\} = \dfrac{\sin\mathbf{y}}{1 - R(A).(1-\sin\mathbf{y})} \cdot N_\parallel \{\underline{\mathbf{P}}(A)\} \\[2ex] \textit{Avec} \quad R(A) = \dfrac{1}{(1-\sin\mathbf{y})} \cdot \left[ 1 - \dfrac{N_\parallel \left\{\sum_A \underline{\mathbf{p}}(m/n)\right\}}{\sum_A N_\parallel \{\underline{\mathbf{p}}(m/n)\}} \right] \\[4ex] \textit{Le taux de réalimentation interne } R(A) \textit{ vérifie: } 0 \le R \le 1 \end{cases}$$

Par rapport à la forme condensée de l'équation de dissipation du contact élémentaire (1.8), observons que, vis à vis du processus de dissipation d'énergie par friction au sein de l'amas granulaire en mouvement, cette grandeur $R$ joue le rôle d'une sorte de ***fonction d'état***, qui porte l'effet de population.

En effet, par sa définition (2.6), cette fonction est irréductiblement liée aux aspects statistiques du mouvement dans l'amas granulaire : si la population est réduite à 1 seul contact en mouvement, alors $R$ devient nul, et donc l'équation de dissipation (2.8) se réduit à l'équation de dissipation du contact élémentaire (1.6), l'effet de population disparaît alors de l'équation de dissipation.

Nous verrons plus loin que:

-cette fonction $R$ croît avec le degré de désordre dans la population des contacts en mouvement; elle peut aussi être interprétée comme une ***mesure*** – significative pour ce processus dissipatif particulier- ***du degré de désordre*** dans le mouvement du matériau;

-son minimum $R=0$ est associé, de façon biunivoque, avec des structures ordonnées dans l'arrangement statistique des mouvements dans l'amas granulaire;

-ces structures ordonnées qui minimisent la fonction $R$, sont en nombre limité, et constituent les ***structures de dissipation minimale***, pour le processus dissipatif analysé ici.

D'autres propriétés de fonction d'état peuvent être dégagées, remarquons en particulier la suivante:

- désignons par $\mathbf{\textit{E}}(A)$ la fonction $\mathbf{\textit{E}}(A) = R(A).\sum_A N_\parallel \{\underline{\mathbf{p}}(m/n)\}$, qui est toujours positive ou nulle au cours du mouvement (elle s'annule avec $R$) ;

- si nous formons un amas par réunion de plusieurs sous-amas granulaires $A \equiv A_1 \cup A_2 .... \cup A_n$, la fonction précédente vérifie les propriétés (montrées en Annexe 2-1):

    a) $\boldsymbol{E}(A_1 \cup A_2 .... \cup A_n) \geq \boldsymbol{E}(A_1) + \boldsymbol{E}(A_2) + .... + \boldsymbol{E}(A_n)$

    b) l'égalité n'est assurée que lorsque les réalimentations entre deux sous-amas de la partition sont toutes nulles: en l'absence d'interaction entre les sous-amas, la fonction $\boldsymbol{E}(A)$ devient *additive*;

    c) lorsque en plus de l'absence d'interaction entre sous-amas, la fonction $R$ est nulle sur chacun des sous-amas, alors tous les termes de l'inégalité a), et donc la fonction $\boldsymbol{E}(A)$, *s'annulent*. Nous verrons un peu plus loin que cette situation est celle de la dissipation minimale.

-la fonction $\boldsymbol{E}(A)$ présente donc des propriétés analogues à celles de certaines fonctions d'état thermodynamiques.

## *2.5 EXTREMUM DE LA DISSIPATION*

Considérons l'amas granulaire comme un système échangeant de l'énergie avec l'extérieur, de par la puissance développée par les efforts extérieurs, qui se trouve dans notre analyse, égale au signe près à la puissance des efforts intérieurs (hypothèses de quasi-équilibre).

Dans son bilan interne, notre système reçoit de l'énergie (puissance reçue $\mathbf{P}^+(A)$) et en rend (puissance rendue $\mathbf{P}^-(A)$); la somme algébrique de ces deux termes est la puissance des efforts intérieurs, dissipée par friction.

Définissons l'intensité de cette dissipation interne, par un "Taux de Dissipation" $T_D$, que nous rapporterons à la puissance reçue $\mathbf{P}^+(A)$:

$$T_D = \frac{Tr\{\underline{\mathbf{P}}(A)\}}{\mathbf{P}^+(A)} \qquad (2.9)$$

On trouve, à partir des relations précédentes (1.6) et (2.6):

$$T_D = \frac{2\sin\boldsymbol{y}}{(1+\sin\boldsymbol{y}) - R(1-\sin\boldsymbol{y})} \qquad (2.10)$$

$R$ étant compris entre 0 et 1, on voit que le taux de dissipation $T_D$ est fonction croissante du taux de réalimentation interne $R$, et qui varie entre deux extrêmes, figure 2-3:

$$\begin{cases} \text{dissipation Maximale} \quad T_D = 1 \text{ , pour } R = 1 \\ \text{dissipation minimale} \quad T_D = \dfrac{2\sin\psi}{1+\sin\psi} \text{ , pour } R = 0 \end{cases} \quad (2.11)$$

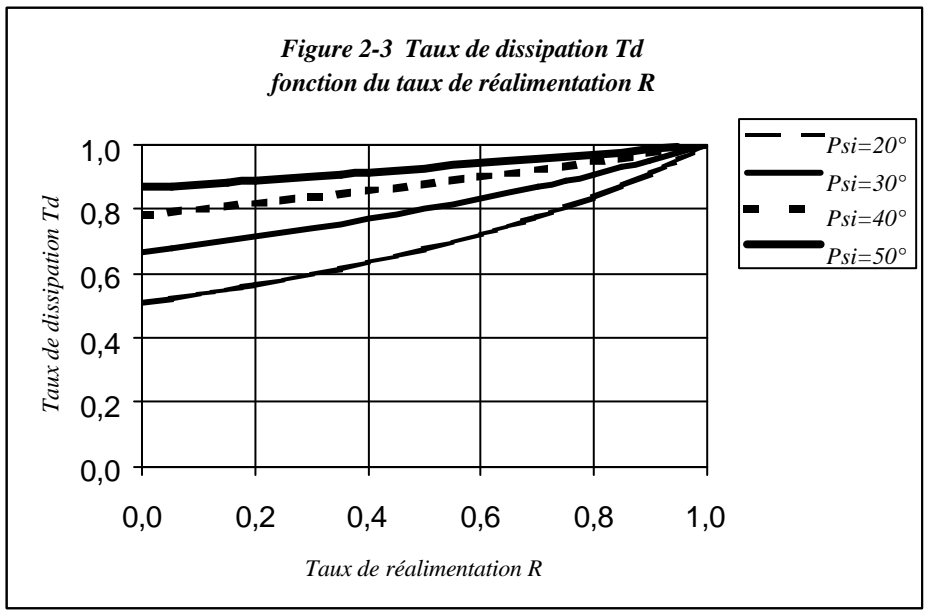

*Figure 2-3 Taux de dissipation Td fonction du taux de réalimentation R*

Si nous calculons ce Taux de Dissipation pour le contact élémentaire isolé, comme dans ce cas $R = 0$, nous trouverons invariablement la valeur minimale.
Lors du passage à l'amas granulaire, l'*effet de population*, dû à la réalimentation interne et représenté par la fonction $R$, substitue à cette valeur unique du Taux de Dissipation, toute une plage de valeurs possibles, qui correspond à l'ensemble des états statistiques dans cette population, admissibles par l'équation de la dissipation (2.8).

Montrons maintenant que les valeurs de ce taux de dissipation dans l'amas granulaire, sont aussi liées au degré de désordre dans la distribution des orientations des contacts en mouvement, au sein de l'amas granulaire.

## *2.6 DISSIPATION MINIMALE ET STRUCTURES ORDONNÉES*

La dissipation minimale correspond à la réalimentation minimale, d'où, d'après les relations (2.11) et (2.7):

$$\begin{cases} R = 0 \\ \mathbf{P}^+(A) = \sum_{m<n\leq N} \mathbf{p}^+(m/n) \\ \mathbf{P}^-(A) = \sum_{m<n\leq N} \mathbf{p}^-(m/n) \end{cases} \quad (2.12)$$

$\mathbf{P}^+(A)$ et $\mathbf{P}^-(A)$ vérifient donc la même relation que les $\mathbf{p}^+(m/n)$ et les $\mathbf{p}^-(m/n)$:
(2.13)
$$Tr\{\underline{\mathbf{P}}(A)\} = \mathbf{P}^+(A) + \mathbf{P}^-(A) = \sin y \{\mathbf{P}^+(A) - \mathbf{P}^-(A)\} = \sin y \cdot N_\parallel \{\underline{\mathbf{P}}(A)\}$$
L'équation (2.13) est donc la formulation tensorielle de la dissipation minimale d'énergie, pour l'amas granulaire.

L'absence de toute réalimentation entraine, d'après (2.6):
$$N_\parallel \left\{ \sum_{m<n\leq N} \underline{\mathbf{p}}(m/n) \right\} = \sum_{m<n\leq N} N_\parallel \{\underline{\mathbf{p}}(m/n)\} \qquad (2.14)$$
Or les propriétés de la norme $N_\parallel$ (voir Annexe 1.2) font que cette égalité n'est possible que si:

a) soit toutes les directions propres portant les $\mathbf{p}^+(m/n)$ sont identiques, et coïncident avec la direction portant $\mathbf{P}^+(A)$, qui est unique dans ce cas, situation que l'on désignera par "Mode I";

b) soit toutes les directions propres portant les $\mathbf{p}^-(m/n)$ sont identiques, et coïncident avec la direction portant $\mathbf{P}^-(A)$, qui est unique dans ce cas, situation que l'on désignera par "Mode II";

c) soit les deux conditions précédentes sont réalisées simultanément, ce qui impose qu'une des composantes de $\mathbf{P}(A)$ soit nulle, situation que l'on désignera par "Mode frontière en déformation plane".

Ces conditions imposent un certain ordre dans la distribution des orientations des tenseurs $\underline{\mathbf{p}}(m/n)$.

Or, l'orientation de chacun des tenseurs $\underline{\mathbf{p}}(m/n)$, donnée par ses directions propres, définit aussi l'orientation des plans moyens des contacts en mouvement correspondants (section 1, Fig 1-2).

On voit ici apparaître le lien entre dissipation minimale, et structure ordonnée dans la distribution des orientations des contacts en mouvement.

### *2.6.1 Mode I (Figure 2-4)*
La valeur propre positive de $\underline{\mathbf{P}}$, unique en son genre, est la somme de toutes les valeurs propres positives des tenseurs élémentaires $\underline{\mathbf{p}}(m/n)$, les directions propres portant les $\mathbf{p}^+(m/n)$ sont identiques, cette direction propre commune est aussi celle de $\mathbf{P}^+$.

Les directions propres élémentaires portant les $\mathbf{p}^-(m/n)$ sont donc distribuées dans le plan perpendiculaire, contenant aussi les deux autres directions propres de $\underline{\mathbf{P}}$.

La propriété de *conservation par rotation du repère* (section 1-2-3), assure que, quelle que soit la distribution des $\mathbf{p}^-(m/n)$ dans ce plan, on en retrouvera toujours l'intégralité dans la somme des composantes sur les axes propres de $\underline{\mathbf{P}}$ dans ce plan.

   Dans le repère propre, la distribution des orientations des plans moyens des contacts en mouvement, correspondant à cette disposition, admet une enveloppe: le double cône de révolution autour de l'axe de $\mathbf{P}^+$, faisant un angle de $\dfrac{p}{4}-\dfrac{y}{2}$ avec cet axe ( généralisation de la Figure1-2, par rotation autour de l'axe 1), et dont les génératrices sont les directions des vecteurs vitesse de glissement.

*Figure 2-4 Structures de dissipation minimale: Mode I*

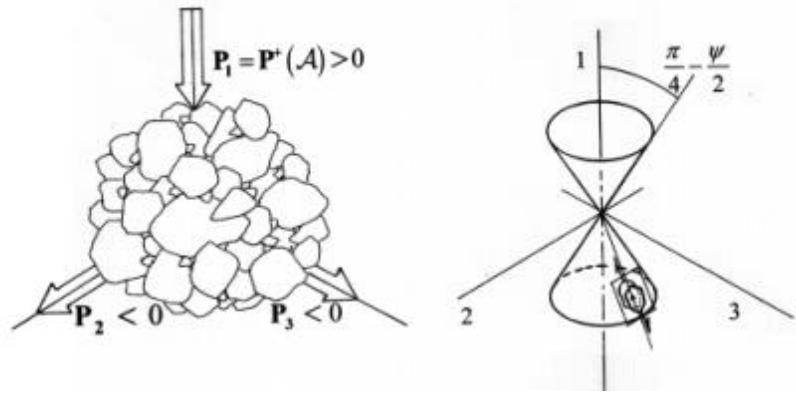

Notons que la configuration de ce double cône est indépendante de la distribution effective des $\mathbf{p}^-(m/n)$ dans le plan orthogonal à l'axe, dont la configuration est comprise entre deux limites: la configuration axissymétrique (distribution uniforme des $\mathbf{p}^-$ dans le plan), et la configuration en déformation plane (tous les $\mathbf{p}^-$ orientés dans une même direction, de même pour tous les axes neutres). Suivant les cas, varie la densité de répartition des plans moyens des contacts, tangents au double cône, mais la géométrie du double cône lui-même demeure fixe pour tout le Mode I.

### 2.6.2  Mode II  (Figure 2-5)

Le raisonnement, symétrique au précédent, s'appuie sur la coïncidence de toutes les directions propres portant les $\mathbf{p}^-(m/n)$, cette direction propre commune est aussi celle de $\mathbf{P}^-$.

Les directions propres portant les $\mathbf{p}^+(m/n)$ sont distribuées dans le plan perpendiculaire, et la distribution des orientations des plans moyens des contacts en mouvement, admet encore une enveloppe: le double cône de révolution autour de l'axe de $\mathbf{P}^-$, faisant un angle de $\dfrac{\pi}{4}+\dfrac{\psi}{2}$ avec cet axe (généralisation de la Figure 1- 2, cette fois par rotation autour de l'axe 3).

*Figure 2-5  Structures de dissipation minimale: Mode II*

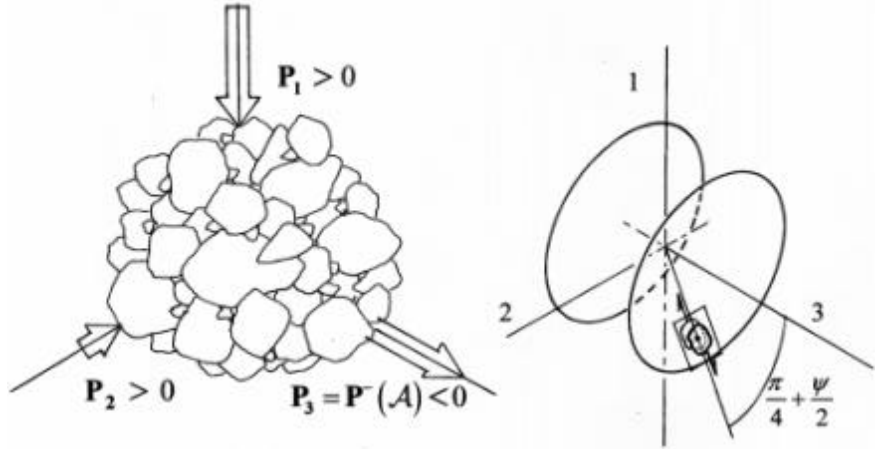

*Figure 2-5  Structures de dissipation minimale: Mode frontière en déformation Plane*

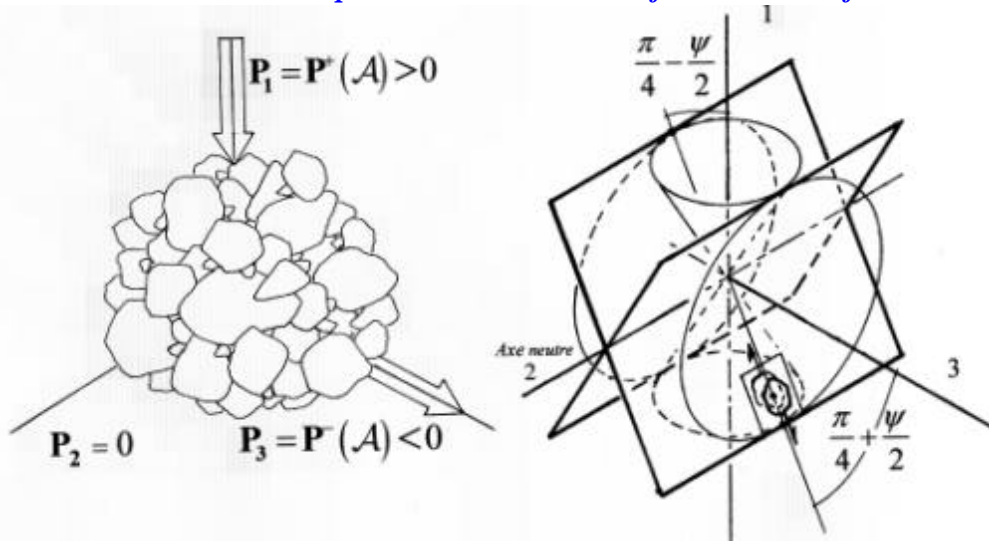

## 2.6.3 Mode frontière en déformation plane  (Figure 2-6)

Les Modes I et II précédents, présentent une frontière commune, lorsque l'une des valeurs propres de $\underline{\mathbf{P}}$ est nulle.

A la différence des situations avec réalimentation interne (section 2.4), ici $\underline{\mathbf{P}}$ ne peut présenter qu'une seule valeur propre nulle, montrons que, de plus, tous les axes neutres des tenseurs élémentaires $\underline{\mathbf{p}}(m/n)$ sont alors identiques à la direction portant cette valeur propre nulle de $\underline{\mathbf{P}}$.

En effet, dans le Mode I, une même direction propre porte tous les $\mathbf{p}^+(m/n)$ ainsi que la seule composante positive de $\underline{\mathbf{P}}$, alors que dans le Mode II, la situation est symétrique avec les $\mathbf{p}^-(m/n)$.

A la frontière commune des deux modes, les directions propres des tenseurs élémentaires, hors axes neutres, coïncident par genres, il en est donc nécessairement de même pour les axes neutres.

Physiquement ceci signifie qu'il n'y a ni mouvement ni échange d'énergie dans cette direction neutre, à la fois globalement pour l'ensemble de l'amas granulaire, et localement pour chacun des contacts élémentaires en mouvement.

Ce mode frontière correspond donc à une déformation plane, à la fois globale et locale. L'enveloppe des orientations des plans moyens des contacts en mouvement, est la paire de plans tangents communs aux doubles cônes des Modes I et II.

### *2.6.4 Polarisation du milieu granulaire, associée à la dissipation minimale*

Nous verrons plus loin que, sous certaines conditions aux limites, le comportement mécanique expérimental présente une tendance naturelle à tendre vers les Modes de dissipation minimale.

Remarquons que ce qui précède montre que:
- si le comportement tend vers la dissipation minimale, alors la distribution statistique des contacts actifs tend nécessairement vers une des structures ordonnées associées à cette dissipation minimale, et simultanément la fonction *R* tend vers zéro;
- une telle évolution vers la dissipation minimale, s'accompagne donc d'une sorte de ***polarisation progressive*** de la distribution statistique des contacts actifs;
- une mesure du degré de cette polarisation est donnée par la fonction 1-*R*, ce qui souligne encore le rôle de fonction d'état joué par *R*.

### *2.6.5 Lignes de glissement des équilibres de Rankine*

Dans le mode frontière en déformation plane, en coupe perpendiculaire à l'axe neutre, les traces des orientations des contacts en glissements, sont donc des lignes inclinées à $\frac{p}{4} - \frac{y}{2}$ par rapport à l'axe "actif" correspondant à la puissance reçue $\mathbf{P}^+$, et inclinées à $\frac{p}{4} + \frac{y}{2}$ sur l'axe "passif" de la puissance rendue $\mathbf{P}^-$.

Cette configuration est connue depuis longtemps en géomécanique sous le nom de **_lignes de glissement de Rankine_** [2]. Dans la schématisation en milieu continu de ces matériaux, ces lignes font partie de la théorie de la rupture dite "des équilibres limite", toujours utilisée dans la pratique, du fait de son efficacité largement prouvée.

La justification de ces lignes de glissement de Rankine, à partir de la physique statistique à l'échelle de la microstructure, constitue une des vérifications essentielles de l'approche exposée ici.

La réalité matérielle de ces lignes de glissement, d'orientation bien définie, suggère par ailleurs qu'au voisinage des équilibres limite, la mécanique de ces matériaux suit un principe de moindre énergie: celui de la dissipation minimale.

### *2.6.6 Exclusivité des Modes*

Notons qu'en tridimensionnel vrai, les Modes I et II s'excluent l'un l'autre: un mouvement de dissipation minimale dans l'amas granulaire ne peut pas être du Mode I sur une partie de l'amas, et du Mode II sur l'autre partie (par contre, une partie de l'amas peut se trouver en déformation plane à la frontière du Mode, et une autre partie peut demeurer inactive, *section 2-8-1*).

### *2.6.7 Similitude interne-Signature de Mode*

Le Mode frontière en déformation plane présente une remarquable propriété de ***similitude interne***: les tenseurs de puissance de l'amas granulaire entier et de tout sous-amas, ont la même orientation, sont homothétiques, et vérifient les mêmes équations (les deux relations (1.4)) que chacun des contacts élémentaires.

Cette similitude interne n'est plus complète dans les deux autres Modes, l'existence d'un axe neutre et son unicité sur tout l'amas, ainsi que la deuxième des relations (1.4), ne sont plus assurées; par contre toutes les autres propriétés sont conservées.

Remarquons parmi celles-ci la propriété de ***signature de mode***, utile par la suite: dans un amas en dissipation minimale, rapporté au repère propre du tenseur $\mathbf{P}(A)$, la diagonale du tenseur de puissance de tout sous-amas conserve une signature fixe (*Si ce n'était pas le cas, il y aurait au moins un $\underline{\mathbf{p}}(m/n)$ qui ne serait pas dans l'ordre*).

### *2.7 DISSIPATION MAXIMALE ET STRUCTURES DESORDONNEES*

La dissipation maximale correspond à la réalimentation maximale:

$$\begin{cases} R = 1 \\ \mathbf{P}^-(A) = 0 \end{cases} \qquad (2.15)$$

Toutes les valeurs propres de $\underline{\mathbf{P}}(A)$ sont alors positives ou nulles: l'amas granulaire reçoit de l'énergie, au sens large, dans toutes les directions, il ne rend d'énergie dans aucune direction.

On a vu que la réalimentation interne est rendue possible par certaines configurations des orientations mutuelles des tenseurs de puissance élémentaires (section 2.3.2). Si la distribution des orientations des tenseurs de puissance élémentaires $\underline{\mathbf{p}}(m/n)$ est suffisamment désordonnée, on y trouvera toujours lieu à réalimentation, qui se traduit ensuite sur la sommation qui forme le tenseur global $\underline{\mathbf{P}}(A)$. On voit donc apparaître le lien entre réalimentation interne, et degré de désordre dans la distribution des orientations des contacts en mouvement.

A la différence de la dissipation minimale, nous constaterons que les solutions admissibles sont nombreuses, non exclusives, et combinables entre elles. Elles incluent un sous-ensemble remarquable de solutions particulières, détaillé en Annexe 2-2, constituées par la combinaison/juxtaposition de mouvements suivant les deux Modes de dissipation minimale, convenablement dosés.

Parmi les distributions théoriquement admissibles des orientations des tenseurs élémentaires $\underline{\mathbf{p}}(m/n)$, vérifiant (2.15), certaines , schématisées Figure 2-7, sont particulièrement remarquables (détails en Annexe 2-2, la liste n'est pas exhaustive):

a) <u>monoaxiales-</u> le tenseur $\underline{\mathbf{P}}(A)$ résultant n'a qu'une seule valeur propre positive, dont la direction est prise comme axe de référence:
   -distribution axissymétrique simple,
   -distribution axissymétrique double, obtenue par composition/juxtaposition de mouvements suivant les Modes de dissipation minimale, de même axe , mais *contraposés*;
   - distribution symétrique double, obtenue par composition/juxtaposition de 2 mouvements suivant le Mode II de dissipation minimale, d'axes orthogonaux à l'axe de référence de $\underline{\mathbf{P}}(A)$, et orthogonaux entre eux (non représentée sur la figure, détails en Annexe 2-2);
- autre solutions, incluant toutes combinaisons linéaires à coefficients positifs des solutions précédentes.

b) <u>biaxiales -</u> le tenseur $\underline{\mathbf{P}}(A)$ résultant a deux valeurs propres positives:
   -distributions obtenues par composition de deux distributions monoaxiales;
   -distribution double, obtenue par composition/juxtaposition de mouvements *contraposés* suivant les Modes de dissipation minimale;
   -distribution orthotrope, où tous les axes neutres des tenseurs élémentaires sont parallèles;

-autres solutions, incluant toutes combinaisons linéaires (à coefficients positifs) des précédentes.

*Figure 2-7 Quelques solutions particulières de la dissipation Maximale*

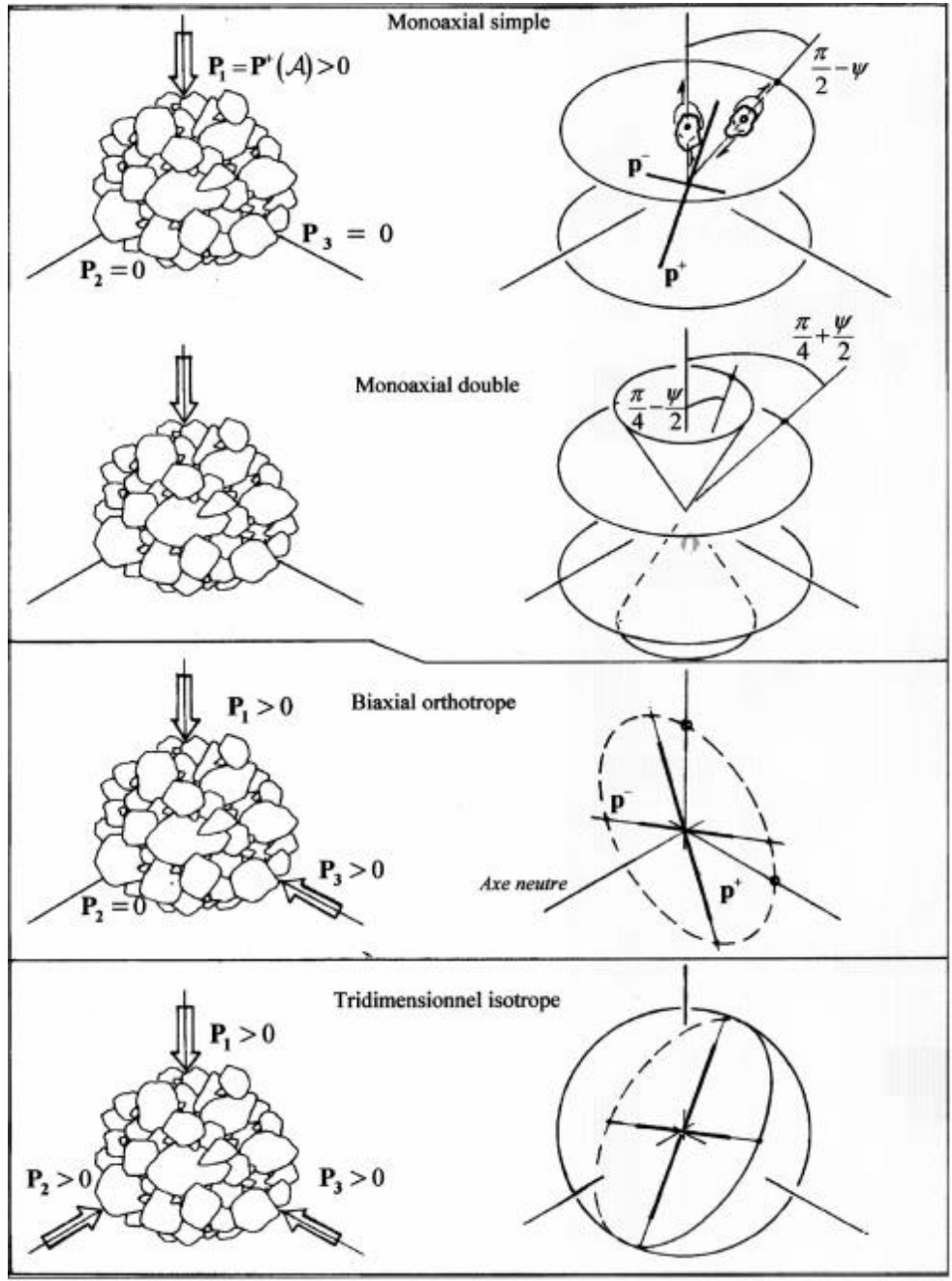

c) <u>tridimensionnelles</u> - le tenseur $\underline{\mathbf{P}}(A)$ résultant a trois composantes positives:

-distributions obtenues à partir des précédentes, par composition de trois distributions monoaxiales, de trois distributions biaxiales, ou d'une biaxiale avec une monoaxiale;

- distribution double, obtenue par composition/juxtaposition de mouvements *contraposés* suivant les Modes de dissipation minimale;
- distribution isotrope, les tenseurs élémentaires étant distribués de façon isotrope;
- autres solutions, incluant toutes combinaisons linéaires (à coefficients positifs) des précédentes.

Enfin, remarquons que toutes combinaisons linéaires, à coefficients positifs, de ces types de distributions, dont les orientations mutuelles seraient non plus fixées suivant des directions propres prédéterminées, mais ***distribuées au hasard dans l'espace***, conduiraient encore à des tenseurs résultants $\underline{\mathbf{P}}(A)$ vérifiant $\mathbf{P}^-(A) = 0$ (*car une somme de tenseurs symétriques à valeurs propres positives, même orientés au hasard, est encore un tenseur symétrique à valeurs propres positives*) ; c'est la propriété de **_convexité_** de l'ensemble des solutions en dissipation Maximale.

La dissipation Maximale correspond donc à une vaste famille de distributions statistiques des contacts en mouvement, sans ordre défini, dont les structures couvrent tous les intermédiaires entre deux types :
- soit des structures désordonnées dans l'amas granulaire: dans toute direction, on trouve "un peu de tout", aussi bien des $\mathbf{p}^+$ que des $\mathbf{p}^-$ ; structures qui se caractérisent donc par une réalimentation interne répartie un peu partout;
- soit des structures composites, avec dans l'amas granulaire, la juxtaposition de sous-amas ordonnés en dissipation minimale, mais entre lesquels les distributions sont différentes; structures qui se caractérisent donc par une réalimentation interne localisée aux frontières entre ces sous-amas.

Remarquons enfin que ces structures désordonnées ou composites, privent les régimes de dissipation Maximale de la propriété de similitude interne: on n'y trouve pas de direction propre *active* commune, ni d'équation de dissipation vérifiée à toutes les échelles.

## *2.8 SOLUTIONS DE L'EQUATION GENERALE DE DISSIPATION 0<R<1*

Avant d'aborder l'analyse des solutions générales de l'équation de dissipation, signalons trois propriétés utiles, liées à la structure de l'équation de dissipation:

-a) certains mouvements admissibles peuvent réaliser une partition de l'amas en sous-amas actifs et sous-amas inactifs;
-b) l'effet de "friction apparente" pour les solutions avec $R \neq 0$;
-c) la décomposition du tenseur de puissance, pour $0<R<1$, en deux composantes coaxiales:
- soit l'une en dissipation minimale, et l'autre en dissipation Maximale;
- soit deux composantes en dissipation minimale .

### *2.8.1 Partition de l'amas en sous-amas actifs et inactifs*

Les grandeurs intervenant dans l'équation générale de la dissipation (2.8) sont toutes relatives aux contacts en mouvement. Donc si l'amas A est formé de la réunion d'un sous-amas actif $A_1$ et d'un sous-amas inactif (c'est à dire une "zone morte" sans aucun contact actif, dont le mouvement est donc celui d'un corps solide), le tenseur de l'amas $\underline{\mathbf{P}}(A)$ est en fait celui du sous-amas actif $A_1$, lequel vérifie donc l'équation (2.8) dans les mêmes conditions que $\underline{\mathbf{P}}(A)$.

Symétriquement, on peut réunir un amas actif avec un amas inactif, le tenseur de l'amas global demeure égal au tenseur de la partie active , et donc vérifie l'équation (2.8) dans les mêmes conditions.

Enfin le nombre de sous-amas actifs et inactifs n'est pas limité**:** un mouvement vérifiant l'équation générale de dissipation (2.8) peut être formé par la juxtaposition d'un grand nombre de "cellules actives" et de "cellules inactives" au sein de l'amas granulaire, ces "cellules " pouvant elles-mêmes être ordonnées en structures organisées .

Dans cette famille de solutions, se trouvent donc:
- des mouvements où de grandes masses qui se comportent comme des corps solides, sont séparées par des zones qui concentrent les déformations, effets bien connus en géomécanique sous les désignations de "zones de localisations des déformations", "zones de cisaillement", ou encore "surfaces de glissement";
- des mouvements où une masse granulaire en déformation, entraîne dans son mouvement des "grumeaux" qui se comportent comme des corps solides, effets parfois observés dans de grands glissements de terrain.

### *2.8.2 Friction apparente*

Pour les solutions tensorielles de (2.8) dans une situation avec $0<R<1$, tout se passe apparemment comme pour la dissipation minimale, mais avec un angle de friction *apparent* $\mathbf{y}^*$, qui s'exprime en fonction de $R$ et de l'angle de friction physique grain sur grain $\mathbf{y}$, par:

$$\sin \mathbf{y}^* = \frac{\sin \mathbf{y}}{1-(1-\sin \mathbf{y})R} \qquad (2.16)$$

La Figure 2-8 en donne une représentation graphique.

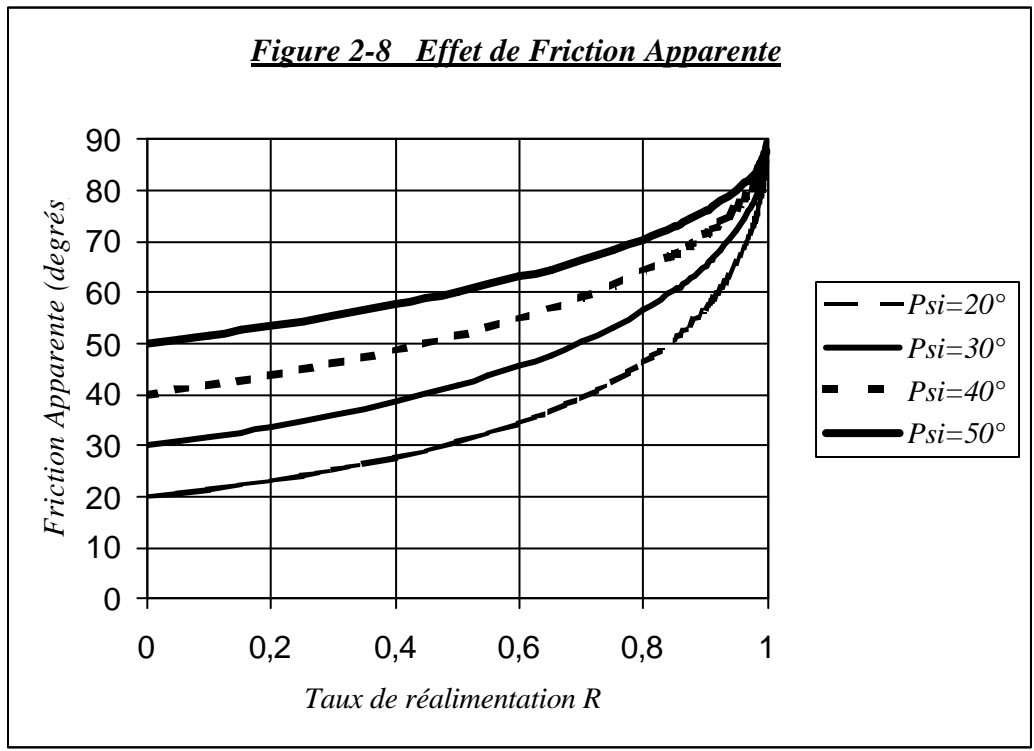

Toutefois, à la différence de $\mathbf{y}$, $\mathbf{y}^*$ n'est plus une constante physique intrinsèque et permanente du matériau, car, par la variable *R*, $\mathbf{y}^*$ dépend de l'arrangement momentané de la population des contacts en mouvement.

Par ailleurs, comme $R \neq 0$, la distribution des orientations des contacts en mouvement n'est plus complètement polarisée comme dans les véritables situations de dissipation minimale.

### 2.8.3 Décomposition du tenseur de puissance

Dans une situation avec 0<*R*<1:

- le tenseur de puissance $\underline{\mathbf{P}}(A)$ présente une puissance rendue non nulle $\mathbf{P}^-(A) \neq 0$ ;
- la puissance reçue $\mathbf{P}^+(A)$ est supérieure au minimum qui serait requis dans un mode de dissipation minimale $\mathbf{P}^+(A) > -\left(\frac{1+\sin\mathbf{y}}{1-\sin\mathbf{y}}\right)\mathbf{P}^-(A)$.

Ces particularités permettent de considérer dans tous les cas, que le tenseur $\underline{\mathbf{P}}(A)$, avec 0<*R*<1, est la somme de deux tenseurs coaxiaux, correspondant à la combinaison/juxtaposition de deux composantes de mouvement:

- une composante en dissipation minimale, dont le tenseur de puissance

$\underline{\mathbf{P}}_m(A)$ est tel que:
$$\begin{cases} \mathbf{P}_m^{+}(A) = -\left(\frac{1+\sin y}{1-\sin y}\right)\mathbf{P}_m^{-}(A) \\ \mathbf{P}_m^{-}(A) = \mathbf{P}^{-}(A) \end{cases}$$

- une composante en dissipation Maximale, dont le tenseur de puissance $\underline{\mathbf{P}}_M(A)$ est tel que:

$$\begin{cases} \mathbf{P}_M^{+}(A) = \mathbf{P}^{+}(A) - \mathbf{P}_m^{+}(A) \\ \mathbf{P}_M^{-}(A) = 0 \end{cases}$$

Pour faire réapparaître la deuxième forme remarquable de décomposition du tenseur de puissance, mais ici avec $0 < R < 1$, il convient d'abord de remarquer quelques particularités géométriques.

*Figure 2-9    Ensemble des solutions de l'équation générale de la dissipation.*
*Représentation géométrique.*

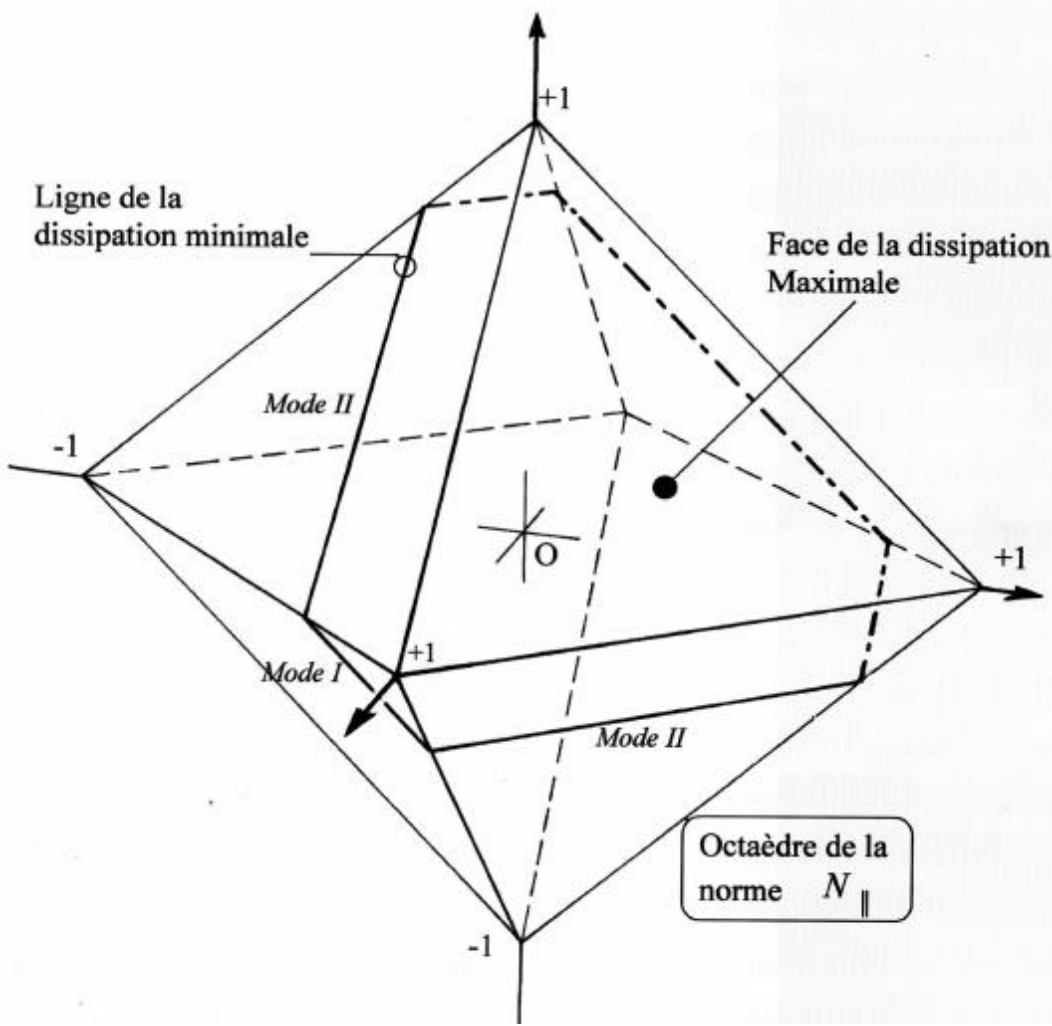

## *2.8.4 Représentation géométrique des solutions*

Une représentation de l'ensemble des solutions de (2.8), dans le repère propre, est donnée Figure 2-9, en normalisant les tenseurs $\underline{\mathbf{P}}$, par la norme $N_\|$ ($N_\|\{\underline{\mathbf{P}}\} = |\mathbf{P}_1| + |\mathbf{P}_2| + |\mathbf{P}_3|$), dont la boule unité est l'octaèdre de la figure:

- l'ensemble des solutions en dissipation minimale (cas de *R*=0, sections 2-5 et 2-6), est une ligne fermée, intersection de l'octaèdre avec le plan $Tr\{\underline{\mathbf{P}}\} = \sin\mathbf{y}$ ; cette ligne est un hexagone, symétrique mais irrégulier, dont les petits côtés correspondent au Mode I, et les grands côtés au Mode II, les sommets représentant les Modes frontières en déformation plane;
- l'ensemble des solutions en dissipation Maximale (cas *R*=1, sections 2-5 et 2-7), est la face (+,+,+) de l'octaèdre, incluant les arêtes et sommets qui la bordent;
- l'ensemble des solutions intermédiaires 0<*R*<1 (section 2-8), est la partie de l'octaèdre comprise entre la ligne de la dissipation minimale et la face de la dissipation Maximale;
- toute la partie de l'octaèdre située en dessous du plan $Tr\{\underline{\mathbf{P}}\} = \sin\mathbf{y}$ (en dessous de la ligne de dissipation minimale) est **<u>exclue</u>**.

Cette représentation géométrique illustre en particulier les différences essentielles entre les ensembles de solutions en dissipation minimale d'une part, et d'autre part en dissipation Maximale:

- l'ensemble des solutions en dissipation minimale est simplement *connexe*, convexe par segments, ce qui traduit l'exclusivité des Modes: les Modes de dissipation minimale ne sont pas combinables entre eux *(mais présentent la similitude interne)*;
- l'ensemble des solutions en dissipation Maximale est *convexe*: les régimes de dissipation Maximale sont parfaitement "miscibles " entre eux *(mais ne présentent pas de similitude interne)*;
- la dissipation Maximale peut se manifester en monoaxial, biaxial, ou tridimensionnel, tandis que la dissipation minimale ne peut se manifester qu'en biaxial ou tridimensionnel: il n'y a pas de mode de dissipation minimale monoaxial;
- la décomposition donnée plus haut en section 2.8.3, se retrouve dans la représentation géométrique: l'ensemble des solutions de l'équation générale de dissipation (2.8) avec $0 \leq R \leq 1$, est un convexe, dont les solutions de dissipation minimale (pour *R*=0), forment le bord.

Dans cette représentation géométrique, remarquons que l'ensemble des résultats qui précèdent permettent de déduire que toute combinaison linéaire *à coefficients positifs* de 2 solutions en dissipation minimale, est encore une solution de l'équation générale de dissipation**:** si $m'$ et $m''$ sont deux points figuratifs de solutions en dissipation minimale sur l'octaèdre, Figure 2-10, les points figuratifs de l'ensemble des combinaisons linéaires à coefficients positifs , sont sur l'intersection de l'octaèdre avec le secteur de plan défini par les deux vecteurs $Om'$ et $Om''$, ligne $m'Mm''$ sur la figure.

En un point *M* donné sur l'octaèdre, tel que $R \neq 0$, passent donc une infinité de lignes du type $m'Mm''$, résultant de l'intersection de l'octaèdre avec le faisceau de plans d'axe $OM$ ; toute solution avec $R \neq 0$, peut donc être décomposée d'une infinité de façons sous forme d'une combinaison linéaire de solutions en dissipation minimale.

Dans ce faisceau, se distinguent les ensembles de solutions particulières correspondant à une combinaison de Modes I et II de ***même axe,*** et ***contraposés*** ( points figuratifs $m_I$ et $m_{II}$ sur deux faces diamétralement opposées de l'octaèdre):

- si *R*=1, point *M* sur la face (+,+,+), il y a 3 ensembles de solutions de ce type, données géométriquement par les plans passant par *M*, le centre de l'octaèdre, et le voisinage d'un des sommets de cette face ;

***Figure 2-10 Solutions de l'équation de dissipation pour*** $R \neq 0$
*Décomposition en combinaison de modes de dissipation minimale.*

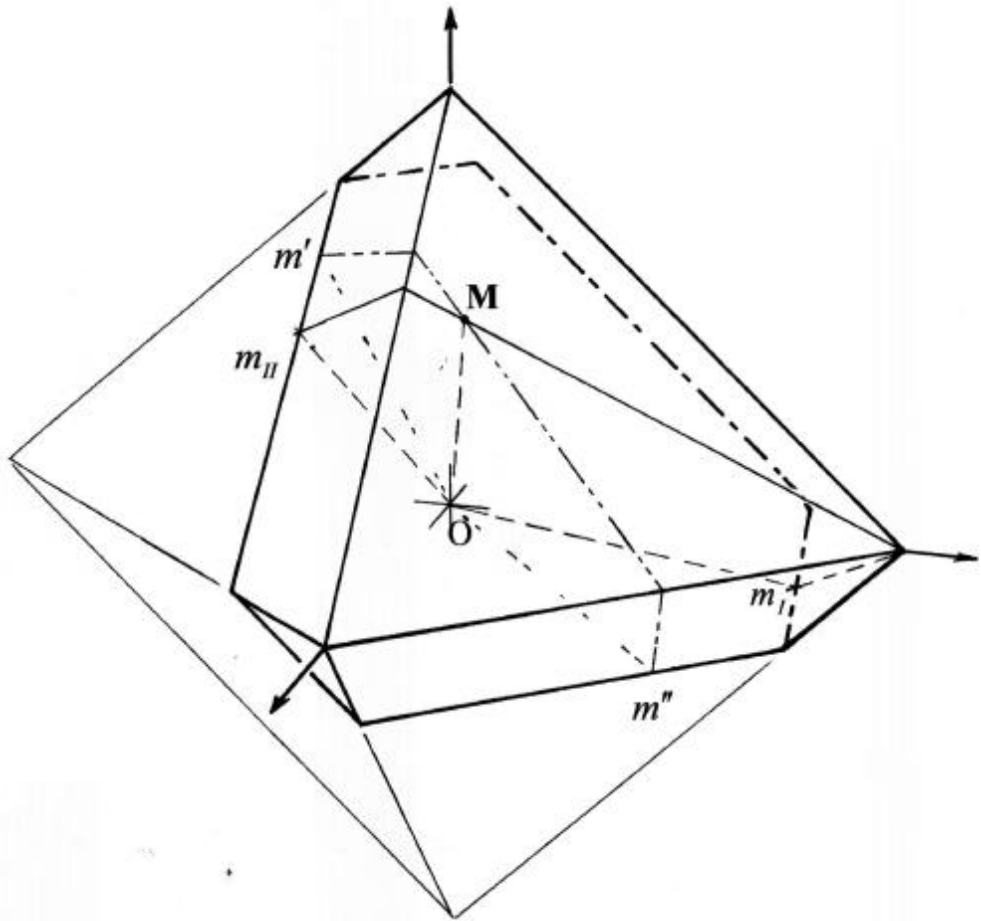

- si $0<R<1$, point *M* à l'extérieur de la face (+,+,+), il y a encore au moins un ensemble de solutions de ce type, donnée par les plans passant par *M*, le centre de l'octaèdre, et le voisinage du sommet de la face (+,+,+) le plus éloigné de *M*.

Suivant les proportions des deux composantes, on décrira ainsi tout l'ensemble des solutions tensorielles de l'équation générale de dissipation.

L'Annexe 2-2-B détaille ces proportions, ainsi que les *participations de chacune des composantes à la dissipation d'ensemble*, pour les cas de dissipation Maximale monoaxiale, biaxiale, et tridimensionnel.

De manière analogue, on peut construire un ensemble de solutions particulières correspondant à une combinaison de 2 mouvements de même Mode I ou II, mais d'axes orthogonaux (cas de la ligne $m'Mm''$ de la figure).

En conclusion, il apparaît que plusieurs familles de solutions particulières par combinaison /juxtaposition de mouvements en dissipation minimale de Modes I et/ou II, dont celle des mouvements de ***même axe,*** mais ***contraposés***, sont génératrices de tout l'ensemble des solutions tensorielles de l'équation générale de dissipation.

Ceci montre que l'on pourra toujours considérer qu'un mouvement d'ensemble, de type dissipation Maximale (par exemple celui de la "consolidation uniaxiale"):
a) résulte de la composition de deux composantes, chacune séparément en dissipation minimale, et interagissant entre elles par réalimentation interne,
b) permet de déterminer les proportions, ainsi que les participations à la dissipation d'ensemble, de chacune de ces composantes, suivant la famille de solutions particulières choisie.

Cette propriété est à mettre en perspective avec le processus souligné plus loin en sections 3-3 et suivantes, qui fait tendre le mouvement vers la dissipation minimale dès que les conditions aux limites le permettent (par exemple le relâchement progressif du frettage latéral dans un mouvement initié en "consolidation uniaxiale").

## 3 - CONSEQUENCES SUR LES PROPRIETES DU MILIEU CONTINU EQUIVALENT

### *3-1 DISSIPATION D'ENERGIE DANS L' EXPERIMENTATION.*

Dans des publications antérieures portant sur l'interprétation physique du comportement mécanique expérimental de ces matériaux [12],[13], analysé dans une optique de mécanique des milieux continus, l'auteur a montré que, sous des conditions aux limites permettant au matériau de rendre de l'énergie au milieu extérieur **:**
a) le comportement mécanique expérimental vérifie une certaine équation de dissipation qui s'écrit

$$\sum_i \mathbf{s_i} \dot{\mathbf{e}}_i = \sin y \cdot \sum_i |\mathbf{s_i} \dot{\mathbf{e}}_i| \quad (3.1)$$

*où les $\mathbf{s_i}$, $\dot{\mathbf{e}}_i$ sont les valeurs propres des tenseurs eulériens des contraintes et vitesses de déformation $\underline{\mathbf{s}}$ et $\underline{\dot{\mathbf{e}}}$, et $y$ une constante physique intrinsèque du matériau, indépendante des conditions de l'expérience, et en particulier insensible aux variations de compacité, qui s'interprète comme une friction physique moyenne entre les grains;*

b) cette équation de dissipation peut s'écrire en fonction des seules valeurs propres d'un certain « tenseur de puissance » des actions intérieures, symétrique et du second ordre,

défini par
$$\begin{cases} \mathbf{p} = \tfrac{1}{2}\{\underline{\mathbf{s}} \otimes \underline{\dot{\mathbf{e}}} + \underline{\dot{\mathbf{e}}} \otimes \underline{\mathbf{s}}\} \; \text{(contracté)} \\ p_{ij} = \tfrac{1}{2}(s_{ik} \cdot \dot{e}_{kj} + \dot{e}_{il} \cdot s_{lj}) \\ Tr\{\mathbf{p}\} = \sum_{ij} s_{ij} \cdot \dot{e}_{ij} \end{cases}$$, l'équation (3.1) s'écrivant

alors : 
$$\sum_i \mathbf{p_i} = \sin y \cdot \sum_i |\mathbf{p_i}| \quad (3.2)$$

c) de l'équation de dissipation (3.1) ), résultent analytiquement les célèbres relations contraintes-dilatance de Rowe [9][10][11], dans les conditions restrictives considérées par cet auteur (symétrie axiale et déformation plane), *voir Annexe 3* ;

d) lorsque le matériau, soumis à de grandes déformations monotones, tend vers les conditions de déformation à volume stationnaire de « l'état critique », et pour des sollicitations tridimensionnelles, l'enveloppe des situations de moindre dissipation donnée analytiquement par l'équation de dissipation (3.1), est précisément le Critère de rupture de Coulomb (*Annexe 3* ), qui demeure un référence éprouvée en géomécanique, car le Critère de Coulomb est à la base des vérifications de stabilité par la méthode des « équilibres limite » , des méthodes de lignes de glissement, et aussi des méthodes en plasticité (capacité portante des fondations etc..);

e) lorsque le matériau est sollicité par des états de contraintes incluant un petit déviateur, au voisinage d'un état de contraintes isotrope, les déformations correspondantes, décrites par l'équation de dissipation (3.1), résultent toujours en une contraction en volume, correspondant à la notion de « domaine caractéristique » développé sur des bases expérimentales par M.P.Luong [14], *Annexe 3*, et qui correspond aussi à l'effet bien connu de la densification par mouvements alternés, base des techniques de compactage dynamique;

f) la validité expérimentale de l'équation de dissipation (3.1) est établie dans de larges conditions, incluant les conditions axisymétriques (appareil dit « Triaxial »), la déformation plane avec rotation d'axes (appareil « Simple Shear Apparatus ») ou sans rotation d'axes (appareil « Biaxial »), ainsi que pour des sollicitations tridimensionnelles cycliques de grande amplitude (presses tridimensionnelles) [13].

Enfin, on montre *(Annexe 3)* que lorsque le coefficient matériel de friction $\sin y$ tend vers zéro, l'équation ( 3.1) décrit un matériau qui peut être déformé indéfiniment sous contrainte isotrope, sans dissipation d'énergie ni résistance au cisaillement, et dont le volume spécifique devient invariable, ce qui caractérise le comportement du « *fluide parfait incompressible* ». Ce remarquable passage à la mécanique des fluides lorsque le coefficient matériel de friction tend vers zéro, s'explique par le fait que le « matériau » tend alors vers un amas de « billes » plus ou moins convexes et régulières, glissant sans frottement les unes sur les autres tout en demeurant en contact, ce qui constitue un modèle microscopique élémentaire de liquide.

Observons la similitude remarquable, entre cette équation de dissipation (3.1), déduite de l'expérimentation pour décrire le comportement du milieu continu équivalent, et l'équation de dissipation de l'amas granulaire discontinu résultant des sections précédentes 1 et 2.
Cette similitude se prolonge jusqu'à la présence d'un tenseur de puissance des actions intérieures $\mathbf{p}$, qui porte la dualité entre contraintes et vitesses de déformations, tout comme dans l'amas granulaire discontinu, le tenseur de puissance $\underline{P}$ est porteur de la dualité entre forces et vitesses de glissements aux contacts ; de plus, la somme des valeurs absolues au deuxième membre fait retrouver la norme $N_{\|}$ ( *qui n'avait pas éte identifiée comme telle, à l'époque* ).

### *3.2 CORRESPONDANCE DISCONTINU/CONTINU EQUIVALENT- Grandeurs macroscopiques et variables microscopiques-Figure 3-1-*

Notons que sur le plan formel, l'écriture (3.2) de cette équation de dissipation avec les valeurs propres de $\mathbf{p}$, est <u>*identique*</u> à celle du discontinu, en substituant $\mathbf{p}$ à $\underline{P}$.
L'identification énergétique entre l'amas granulaire discontinu et sa représentation par un milieu continu équivalent, mène alors à poser que $\mathbf{p}$ et $\underline{P}$ sont homothétiques, car ces deux tenseurs ont la même *trace* (au coefficient $1/V$ près, la trace des deux tenseurs est la *puissance des efforts intérieurs*, au sens classique du terme), et, représentant tous les deux le même système de flux d'énergie mécanique engendrés par les actions intérieures, ils ont nécéssairement les mêmes directions propres. Il vient alors pour la valeur <u>*moyenne*</u> de $\mathbf{p}$, sur un volume donné $V(\mathrm{A})$ contenant un amas granulaire A :

$$\overline{\mathbf{p}} = \frac{1}{V(\mathrm{A})} \cdot \int_{V(\mathrm{A})} \mathbf{p}\,dv = \frac{1}{V(\mathrm{A})} \cdot \underline{P}(\mathrm{A}) \qquad (3.3)$$

Figure 3-1 **: Milieu continu**
*Equivalence énergétique avec le discontinu*

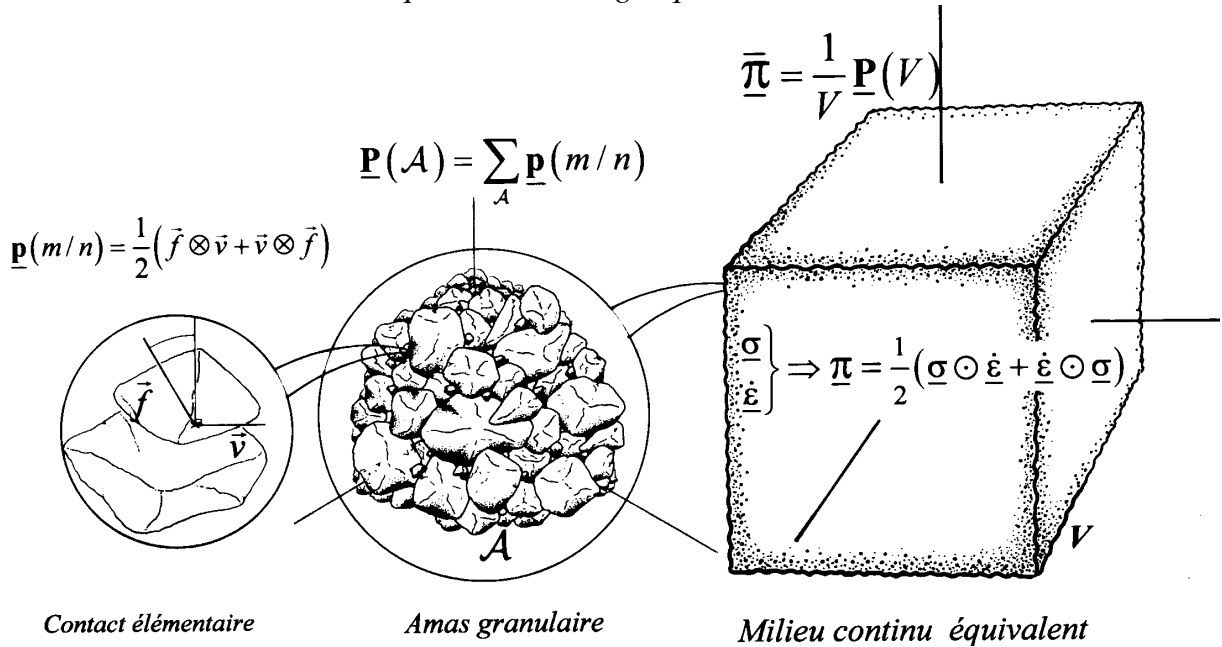

*Contact élémentaire*    *Amas granulaire*    *Milieu continu équivalent*

*Tenseur de puissance des actions intérieures*
Reprenant les définitions des sections 2-3-1 et 1-2-1, nous pouvons maintenant relier directement le tenseur de puissance moyen des actions intérieures du continu équivalent, aux forces de contact et aux vitesses de glissement correspondantes de l'amas granulaire contenu dans le volume V :

$$\overline{\underline{\mathbf{p}}} = \frac{1}{2V} \sum_{n<m\leq N} \left[ \overrightarrow{f(n/m)} \otimes \overrightarrow{v(n/m)} + \overrightarrow{v(n/m)} \otimes \overrightarrow{f(n/m)} \right] \quad (3.4)$$

**Tenseur des Contraintes**

Reprenant les notations de la section 2 sur l'amas granulaire, et considérant un point arbitraire *Gn* rigidement lié au nième granulat, que nous prendrons en son centre de gravité pour fixer les idées, le tenseur des efforts intérieurs de WEBER[19] relie le tenseur des contraintes moyennes, aux forces de contact et à la géométrie interne de l'amas granulaire contenu dans le volume V, il peut être reformulé en une expression symétrique :

$$\overline{\underline{\mathbf{s}}} = \frac{1}{2V} \sum_{n<m\leq N} \left[ \overrightarrow{f(n/m)} \otimes \overrightarrow{GnGm} + \overrightarrow{GnGm} \otimes \overrightarrow{f(n/m)} \right] \quad (3.5)$$

*Tenseur des Vitesses de Déformation*
Par définition, est toujours vérifiée localement l'égalité :
$\underline{\mathbf{p}} = \frac{1}{2}\left[ \underline{\mathbf{s}} \otimes \underline{\dot{\mathbf{e}}} + \underline{\dot{\mathbf{e}}} \otimes \underline{\mathbf{s}} \right]$ (*en produit contracté*)

De plus, dans le cadre de notre hypothèse de base N° 6, l'égalité ci-dessus est aussi vérifiée par les <u>valeurs moyennes</u> sur le volume V (à condition bien sûr qu'il soit assez grand), car l'antisymétrie des covariances assure que le produit symétrique des écarts aux moyennes, est d'espérance mathématique nulle.

Le tenseur des contraintes étant toujours régulier et inversible, du fait de la *condition de non-traction*, sauf peut être près d'une surface libre, l'équation tensorielle résultante est généralement inversible, c'est à dire qu'elle permet de trouver $\overline{\dot{\mathbf{e}}}$, lorsque sont donnés $\overline{\mathbf{p}}, \overline{\mathbf{s}}$.

Toutefois, dans un référentiel quelconque, cette inversion est un peu fastidieuse, elle se simplifie dans le Repère propre des contraintes (référentiel des directions propres en contraintes), car dans ce repère, les composantes vérifient :

$$\overline{p}_{ij} = \tfrac{1}{2} \overline{\dot{e}}_{ij} \left( \overline{\mathbf{s}}_i + \overline{\mathbf{s}}_j \right)$$

Finalement, en résolvant à partir des expressions précédentes, il vient, dans le <u>Repère propre des contraintes</u> :

$$\overline{\dot{e}}_{ij} = \frac{\sum_{n<m\leq N} \left[ f_i(n/m).v_j(n/m) + v_i(n/m).f_j(n/m) \right]}{\sum_{n<m\leq N} \left[ f_i(n/m).GnGm_i + f_j(n/m).GnGm_j \right]} \qquad (3.6)$$

Cette expression, qui est bien de dimension $T^{-1}$, fait intervenir à la fois la cinématique des mouvements, par les vitesses de glissement aux contacts actifs (au numérateur), et la géométrie de l'amas granulaire incluant contacts actifs et contacts inactifs (par les composantes des vecteurs GnGm, au dénominateur), en pondérant ces grandeurs cinématiques ou géométriques, par les forces de contact.

Cette expression se simplifie encore en <u>cas de coaxialité des 3 tenseurs</u> *(on peut montrer que si $\mathbf{s}$ et $\dot{\mathbf{e}}$ sont coaxiaux, alors il en est de même avec $\mathbf{p}$ )*, car dans le repère propre commun, ne subsistent que les termes diagonaux :

$$\overline{\dot{e}}_i = \frac{\sum_{n<m\leq N} \left[ f_i(n/m).v_i(n/m) \right]}{\sum_{n<m\leq N} \left[ f_i(n/m).GnGm_i \right]} \qquad (3.7)$$

La présence des forces de contact, qui pondèrent les termes cinématiques ou géométriques, peut étonner, elle provient directement de la définition physique que nous prenons pour ces vitesses de déformation moyennes : le « quotient » d'une puissance moyenne par une contrainte moyenne. La thermodynamique garantit ( tant que les hypothèses de base du schéma demeurent vérifiées…) que ces déformations moyennes « énergétiques » sont bien représentatives des déformations moyennes au sens usuel du terme.

Par ailleurs, une telle définition « énergétique », peut être rapprochée des méthodes de l'énergie utilisées dans le cadre d'autres types de comportements *(par exemple, en Elasticité, les déformations s'obtiennent comme dérivées partielles de l'énergie élastique, par rapport aux contraintes)*.

Remarquons que ces vitesses de déformation équivalentes s'annulent avec les vitesses de glissements dans le discontinu : c'est une conséquence directe de l'hypothèse de granulats rigides se mouvant les uns par rapport aux autres essentiellement par *glissements*.

En poursuivant l'identification, reste à déterminer si le coefficient matériel de l'équation (3.1) correspond effectivement à $y$, au sens du discontinu, <u>*ce qui impliquerait que l'équation de dissipation « observée » pour le continu est celle de la dissipation minimale*</u>, ou s'il s'agit de la grandeur « apparente » $y^*$ définie en section 2.8.2, avec une valeur non-nulle du taux de réalimentation interne $R$.

## *3.3 TENDANCE EXPERIMENTALE VERS LA DISSIPATION MINIMALE*

Dans l'analyse de l'amas granulaire (Sections 1 et 2), le processus de dissipation général fait intervenir le taux de réalimentation interne $R$, fonction de la distribution statistique des orientations des contacts en mouvement, et donc a priori fonction des conditions initiales.

Par contre, au voisinage des régimes de dissipation minimale, la polarisation de cette distribution *(section 2-6-4)*, associée à la configuration de Mode, maintient nulle la grandeur $R$, et l'équation de dissipation minimale se trouve ainsi "déconnectée" de la mémoire des conditions initiales.

L'interprétation d'expériences sur le milieu continu équivalent, à l'aide de l'équation de dissipation (3.1), peut donc fournir une indication sur le type de mouvement discontinu qui s'y développe (dissipation générale avec $R \neq 0$, ou dissipation minimale avec $R \approx 0$), suivant que les résultats de cette interprétation dépendent ou non des conditions initiales dans le matériau, dont en particulier la compacité.

Dans les expérimentations qui ont fait l'objet d'une interprétation de ce type (voir [13]), il s'avère que, passée une certaine phase d'initiation du mouvement, dont l'amplitude dépend fortement des conditions expérimentales, <u>*les résultats obtenus deviennent indépendants des conditions initiales*</u>.

Cette propriété d'indifférence aux conditions initiales, et donc aussi d'indifférence aux états successifs, est illustrée par la Figure 3-2, obtenue par des mesures de contraintes et déformations sur matériau granulaire calcaire concassé (*donc particulièrement anguleux*), sous contraintes axissymétriques, en variant les conditions initiales de compacité du matériau (proportion de vides par rapport à la matière, ou "indice des vides", variant de 0,75 à 1,17, du plus dense au plus lâche).

Dans ces expériences, on écrase une éprouvette cylindrique de matériau, confiné dans une membrane en caoutchouc, en maintenant constante la contrainte radiale, et en mesurant la contrainte axiale, la déformation axiale, et la variation de volume de l'éprouvette (*essai dit "Triaxial", avec mesure des variations de volume*). Partant d'un état initial de contraintes isotrope, la déformation sous contraintes radiales fixées, induit l'accroissement de la contrainte axiale, et s'accompagne de déformations volumiques significatives, Figure 3-2 a), sur laquelle on peut remarquer qu'au début de l'écrasement, au voisinage de l'état de contraintes isotrope, le matériau exhibe une contraction en volume, qui présente la même pente par rapport aux déformations axiales (Figure 3-2 a), diagramme supérieur), trace expérimentale de la propriété mentionnée plus haut en section 3.1 e).

L'interprétation dont nous parlions plus haut, consiste à tracer les résultats des mesures dans un "diagramme de dilatance", inventé il ya longtemps par Rowe, de coordonnées $\begin{cases} X = 1 - \dfrac{\dot{\varepsilon}_v}{\dot{\varepsilon}_1} \\ Y = \dfrac{\sigma_1}{\sigma_3} \end{cases}$ , en donnant l'indice 1 aux grandeurs axiales.

Dans un tel diagramme, une abscisse inférieure à 1 correspond à un mouvement de contraction en volume, tandis qu'une abscisse supérieure à 1 correspond à un mouvement de dilatance.

Observons que les courbes de résistance et déformation en volume –Figure 3-2 a)- dépendent fortement de la compacité initiale, tandis que dans le diagramme de dilatance – Figure 3-2 b)- les points expérimentaux décrivent une même droite, indépendante de ces conditions initiales.

Or il découle de l'équation de dissipation (3.1) (voir *Annexe 3*), que dans ces conditions expérimentales axisymétriques, les points expérimentaux doivent vérifier :

$\dfrac{\sigma_1}{\sigma_3} = \left(1 - \dfrac{\dot{\varepsilon}_v}{\dot{\varepsilon}_1}\right) \cdot \left(\dfrac{1 + \sin y}{1 - \sin y}\right)$ qui est la célèbre "relation contraintes-dilatance" de Rowe.

Dans le diagramme de dilatance, cette relation est l'équation d'une <u>*droite passant par l'origine*</u>, à condition que $y$ soit bien une constante, et dont la pente vaut précisément $\dfrac{1 + \sin y}{1 - \sin y}$ ( *ce qui constitue un moyen de mesurer cet angle de friction physique $y$ ; les diagrammes de dilatance se révélant par ailleurs un outil efficace de diagnostic du comportement [20],[21]*).

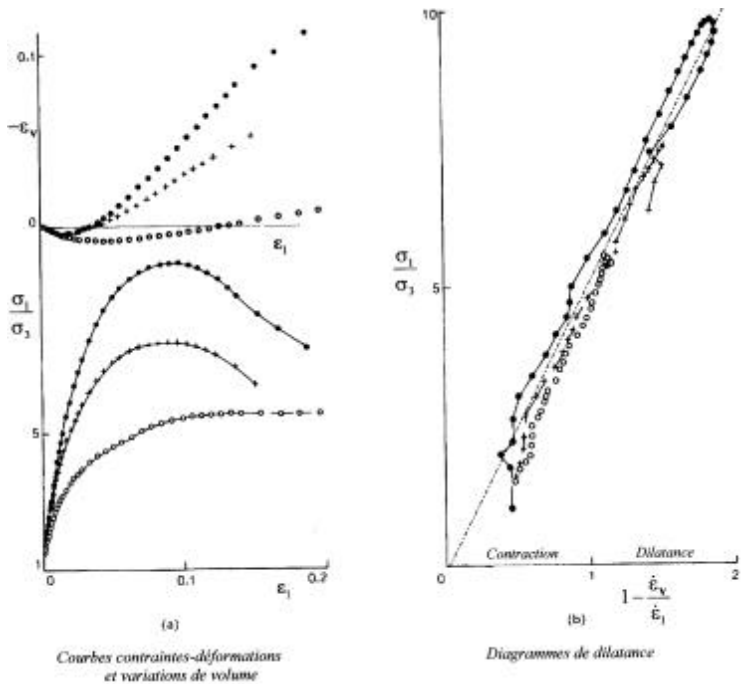

2- ***Figure 3-2:*** *Déformation d'éprouvettes axissymétriques (essais « triaxiaux ») de matériau granulaire calcaire concassé, de compacités initiales différentes:*
· *dense ,*
+ *moyennement dense,*
o *lâche .*

Remarquons enfin que la droite expérimentale dans le diagramme de dilatance demeure insensible à l'amplitude des déformations cumulées, pourtant considérable dans ces expériences**:** les déformations axiales cumulées atteignent 20%, avec des variations de volume spécifique atteignant 12% (soient des distorsions cumulées de l'ordre de 40%, et une augmentation relative de la proportion de vides de l'ordre de 26%).

Il est donc raisonnable de penser que la droite ainsi trouvée correspond à un de nos Modes de dissipation minimale qui s'installe plus ou moins rapidement dans le matériau discontinu. Ici il s'agit du Mode I, et la convergence vers la droite de dissipation minimale est exceptionnellement rapide, par rapport à d'autres types d'expériences [13].
Par ailleurs, l'application de cette méthode à la mesure de la friction moyenne entre grains, pour des matériaux monominéraux, donne des résultats concordants avec d'autres types de mesures [20].

Le comportement expérimental de la Figure 3-2, illustre donc une tendance manifeste vers la dissipation minimale. Cette tendance n'est pas inscrite a priori dans l'équation de la dissipation générale, et résulte donc d'un autre effet physique.
Cette "tendance naturelle" vers la dissipation minimale, dont on pourrait multiplier les preuves expérimentales (voir [13]*),* s'accompagne nécessairement de la décroissance de la fonction *R*, qui tend alors vers 0.
Ceci implique aussi *(voir section 2.6.4),* que la distribution statistique des contacts actifs présente un phénomène de *polarisation spontanée*, dont le développement progressif est étroitement associé à la convergence vers un Mode de dissipation minimale.

Le développement de cette polarisation progressive, associé au mouvement général, n'est vraisemblablement pas instantanné, car la microstructure locale d'un amas granulaire n'a aucune raison d'être à priori compatible avec un mouvement complètement polarisé suivant une direction imposée, et donc l'apparition de cette " tendance naturelle" vers la dissipation minimale requiert certaines conditions aux limites, permettant la polarisation progressive.

Des conditions aux limites plus restrictives (conditions "frettées", rapidement variables, etc.), imposant au matériau de maintenir un taux de réalimentation $R$ significatif en gênant sa polarisation, conduiraient sans aucun doute à d'autres résultats.

Ceci suggère l'hypothèse suivante, qui apparaît pour l'instant comme une hypothèse de travail, vérifiable par l'expérience: *lorsque les conditions aux limites permettent à l'amas granulaire de rendre de l'énergie à l'extérieur, et sont suffisamment régulières, alors le mouvement tend vers un régime de dissipation minimale.*

Par ailleurs, l'analyse de l'amas granulaire *(Section 2)* a montré que:
- toute solution de l'équation générale de dissipation (avec $R \neq 0$) peut être considérée comme somme d'une composante en dissipation minimale (avec $R_1 = 0$), et d'une composante en dissipation Maximale (avec $R_2 = 1$) ;
- les possibilités de mouvement en dissipation minimale sont limitées à l'une des 3 structures ordonnées de la dissipation minimale;
- Les possibilités de mouvement en dissipation Maximale sont illimitées.

La "tendance naturelle" soulignée plus haut, indiquerait donc que, partant de conditions initiales quelconques (avec $R \neq 0$), la composante en dissipation Maximale "s'éteint" progressivement malgré les possibilités de mouvement qui lui sont associées, et le matériau "choisit" l'une des structures ordonnées de la dissipation minimale, dont les conditions de mouvement sont pourtant des plus restrictives.!..

Ce paradoxe apparent, suggère que la dite "tendance naturelle" n'est pas due au hasard. Nous allons voir qu'elle a une raison physique fondamentale, conséquence de la thermodynamique des processus irréversibles.

## 4. SIGNIFICATION PHYSIQUE DE LA TENDANCE VERS LA DISSIPATION MINIMALE

### *4.1 UN PRINCIPE DE MOINDRE ENERGIE*

Les résultats qui précèdent, confirmés par de nombreuses expériences de laboratoires, ainsi que par les pratiques professionnelles dans la conception et la réalisation d'ouvrages de Génie Civil, tournent autour d'un voisinage des conditions de la dissipation minimale, souvent de près, et indiquent la présence d'une sorte de ***principe de moindre énergie* :**

**Lorsque les conditions aux limites le permettent, le matériau tend vers la dissipation minimale.**

## *4.2 UN DEBAT ANCIEN*

L'idée qu'un principe d'extremum est à la base de propriétés physiques de ces matériaux n'est pas nouvelle: elle est déjà présente dans le mémoire de C.A.COULOMB (1773) [1], intitulé d'ailleurs *" Sur une application des règles de Maximis & Minimis à quelques problèmes de Statique,.. » "*.

La combinaison du principe des travaux virtuels avec la méthode de minimisation des forces employée par COULOMB, sur un mécanisme de rupture «*au mouvement commençant* », relierait ses résultats à un principe de moindre énergie.

Un siècle plus tard, L.RAYLEIGH (1873), participant aux développements de la thermodynamique, avait suggéré qu'un « principe de moindre dissipation d'énergie » régissait l'irréversibilité [22].

Plus récemment, dans une analyse de l'amas granulaire restreinte à la symétrie de révolution, sous certaines conditions, ROWE et HORNE (1962-1969 ) [9],[10],[11] avaient bien constaté la nécessité de faire intervenir un principe de minimum, pour établir leur "relation contraintes-dilatance" (*stress-dilatancy formula*) qui est une forme restreinte de notre équation de dissipation minimale tridimensionnelle.

Une dizaine d'années plus tard, cette nécessité d'un principe de minimum a été énergiquement remise en cause par DE JOSSELIN DE JONG (1976)[23], montrant sur un cas très particulier que la relation de ROWE pouvait être établie sans faire intervenir de principe de minimum, dont l'existence en présence de frottements lui paraissait douteuse.

Les développements qui précèdent (sections 1 et 2) montrent que ROWE et DE JOSSELIN DE JONG avaient en fait raison tous les deux, la conclusion finale du second étant toutefois erronée :

- le problème particulier étudié par DE JOSSELIN DE JONG est celui de deux blocs rigides séparés par une surface de glissement, sous contraintes axisymétriques, assimilable au problème du contact élémentaire unique que nous avons analysé en section 1, et pour lequel l'équation de la dissipation minimale est automatiquement vérifiée, car il ne peut y avoir d'*effet de population* dans le cas du contact *unique ;*
- le problème résolu par ROWE est celui de l'amas granulaire soumis à des conditions axissymétriques, cas restreint de l'amas tridimensionnel, pour lequel nous avons vu que l'équation de la dissipation minimale (qui devient ici la"relation contraintes-dilatance" ) n'est vérifiée que lorsque l'*effet de population* dû à la réalimentation interne est *minimal* ;
- le problème résolu par ROWE ne peut donc pas être considéré comme un cas particulier du problème étudié par DE JOSSELIN DE JONG , c'est même précisément l'inverse, et la conclusion de ce dernier n'est donc pas extensible au problème résolu par le premier.

Dans une publication antérieure de l'auteur [13], il avait bien été noté qu'un principe de moindre énergie, à l'échelle du continu, semblait effectivement sous-tendre la réalité du critère de rupture de Coulomb, par rapport à l'équation (3.1).Toutefois cette équation, obtenue à l'époque sur des bases plus intuitives, n'était pas encore clairement identifiée avec la dissipation minimale dans le mouvement de l'amas granulaire discontinu.

## *4.3 LA SIGNIFICATION PHYSIQUE*

Dans l'approche physique présentée ici, ce principe de minimum apparaît lors du passage depuis le comportement microstructural individuel (le contact élémentaire) vers le comportement macrostructural d'ensemble (l'amas granulaire, ou sa représentation par un milieu pseudo-continu équivalent).  Ce principe de minimum met de l'ordre dans « l'effet de population » apporté par la fonction *R,* dont nous avons vu qu'elle présentait des caractères de fonction d'état attachée au désordre dans la distribution statistique des contacts actifs, fonction *R* qui est justement minimale pour la dissipation minimale (sections 2.4 et suivantes)*.*

La *raison d'être* de ce principe de minimum*,* est donc liée à la nature collective, c'est à dire statistique, du processus dissipatif d'ensemble dans l'amas granulaire, elle est donc de *nature thermodynamique.*

En effet, on peut rapprocher le processus physique analysé ici, des résultats sur la thermodynamique des processus dissipatifs au voisinage des conditions d'équilibre, développés depuis une trentaine d'années en particulier par   I.PRIGOGINE et ses collaborateurs [15],[16],[17]:
-dans de telles conditions, ces processus vérifient un *théorème de production minimale d'entropie, qui déduit du deuxième principe que, dans le « domaine linéaire » proche de l'équilibre : «   un système évolue vers un état stationnaire caractérisé par la production d'entropie minimale compatible avec les contraintes imposées au système. Ces contraintes peuvent être déterminées par les conditions aux limites .....l'évolution vers cet état stationnaire signifie l'oubli des conditions initiales particulières. »* (d'après [16],chap. V) ;
-la production d'entropie dans ces processus est liée à l'intensité de la dissipation d'énergie et du développement d'autres irréversibilités, ainsi qu'au développement du désordre statistique.

Dans notre cas de l'amas granulaire, au voisinage des conditions d'équilibre, il est donc naturel de voir apparaître une sorte de convergence vers des solutions de moindre dissipation,  et leur *remarquable coïncidence* avec l'apparition de structures ordonnées dans le "matériau", manifestations concrètes de la production minimale d'entropie, vis à vis du processus dissipatif particulier.

L'existence et la validité de ce principe de moindre dissipation, est de portée générale, nous l'avons vu dans certains résultats, et l'on pourrait en multiplier les preuves expérimentales [13]. Sa correspondance avec le théorème de la production minimale d'entropie, confirme le lien, déjà apparent dans l'analyse de l'amas granulaire, entre la physique statistique et la thermodynamique de l'irréversible d'une part, et d'autre part la présente approche énergétique des matériaux granulaires.
Enfin ce principe de moindre dissipation et sa raison physique, constituent une justification essentielle de la validité de la présente approche énergétique, qui vient parachever les confirmations techniques expérimentales, et celles de la pratique.

Il n'en demeure pas moins surprenant qu'un milieu aussi ordinaire et cahotique, suive d'aussi près une physique aussi « idéale » !

## 5- CONCLUSIONS

L'approche énergétique présentée ici met donc effectivement au jour, un cadre physique qui structure profondément le comportement mécanique, car il est à la source d'un ensemble étendu de propriétés caractéristiques de ces matériaux, utilisées depuis longtemps dans la pratique **:** la densification sous mouvements alternés dans le « domaine caractéristique » (tant que le déviateur des contraintes demeure suffisamment faible), la dilatance sous forts déviateurs suivant les relations de Rowe, le critère de rupture de Coulomb à l'état critique, les lignes de glissement de Rankine....Remarquons qu'au voisinage de la dissipation minimale, tous ces résultats s'obtiennent par une approche physique qui ne dépend que ***d'un seul*** paramètre matériel : la friction moyenne aux contacts entre les grains.

Ce cadre permet aussi de donner une solution au problème du passage entre le discontinu et le continu équivalent, et d'exprimer les grandeurs macroscopiques en fonction des variables microscopiques. La solution donnée ici pour les vitesses de déformation, est originale, à la connaissance de l'auteur. Elle est apportée par l'équivalence énergétique, qui permet au passage de contourner certaines difficultés posées par les méthodes d'"homogénéisation", et assure la transposition au continu équivalent, de propriétés directement héritées du discontinu.

La mise au jour de cette structure physique ouvre aussi une nouvelle brèche dans la vision traditionnelle « purement phénoménologique » des lois de comportement en mécanique des matériaux. Remarquons que cette structure est portée par la notion inédite des *tenseurs de puissance des actions intérieures*, que l'on pourrait abréger en *tenseurs des actions intérieures*, notion qui vient compléter celle, plus traditionnelle, des efforts intérieurs.

Ces actions intérieures sont des grandeurs objectives, tout comme les efforts intérieurs, dont le caractère objectif a été souligné depuis longtemps (par P.Painlevé, vers 1920). C'est précisément ce caractère objectif des actions intérieures, qui confère leur caractère *intrinsèque*, aux propriétés de comportement qui en résultent ; par ailleurs les relations étroites entre ces actions intérieures et l'énergie - grandeur additive- apportent un moyen original de relier le microscopique au macroscopique.

Ce cadre est certes incomplet, car il ne permet pas de déterminer complètement une « loi de comportement ». Toutefois l'ensemble de propriétés qui en résultent, malgré le nombre étonnamment réduit de paramètres matériels en jeu (*un seul…*), indique une structure physique particulièrement claire.

Cette structure provient du jeu combiné de deux effets différents qui contraignent les tenseurs des actions intérieures, tant au niveau local (contact élémentaire) que global (amas granulaire, et continu équivalent):
- *localement*, la loi des actions de contact, ici celle de la friction, qui impose deux relations entre les valeurs propres des tenseurs des actions de contact élémentaires, et qui amène la présence structurante de la norme $N_\parallel$ ;
- *globalement*, la thermodynamique, qui fait tendre le tenseur des actions intérieures de l'amas granulaire vers une solution de dissipation minimale, impliquant la polarisation progressive du matériau, vers une des structures de dissipation minimale, c'est à dire une sorte d'auto-organisation.

Remarquons que l'étape finale de cette convergence vers la dissipation minimale, lorsque toutes les grandeurs deviennent stationnaires, c'est à dire l'"état critique", correspond au critère de rupture de Coulomb, associé à la déformation plane. Cette solution très particulière de la dissipation minimale, car elle est la seule qui réalise complètement la propriété de *similitude interne* (section 2.6.7), semble donc jouer un rôle analogue à celui d'un "attracteur", vis à vis du processus dissipatif particulier.

D'autres lois de contact que la friction, conduiraient sans doute à d'autres types de comportements mécaniques, comme on l'a vu lorsque la friction tend vers zéro, le comportement de notre « milieu granulaire » tendant alors vers le « fluide parfait incompressible » (section 3.1). Ce passage continu vers la mécanique des fluides confirme le rôle-clé des actions intérieures dans la physique du comportement, d'une part, et d'autre part apporte une nouvelle nuance à la distinction traditionnelle entre le *solide* et le *liquide*. On peut envisager d'étendre la méthode des tenseurs des actions intérieures, défrichée dans ces pages, à d'autres collections d'objets matériels, grands ou petits, interagissant entre eux par contacts mutuels, différents de nos contacts frictionnels (contacts élastiques, avec endommagement, etc).

Enfin nous nous sommes attachés ici aux seules actions intérieures, car ce sont elles qui structurent le comportement mécanique intrinsèque de ces matériaux. On pourrait aussi envisager d'étendre la méthode des tenseurs de puissance exposée ici:
- aux actions extérieures (*en utilisant le produit tensoriel symétrique des forces extérieures par les vitesses des points matériels auxquels elles s'appliquent*), ce qui pourrait donner un nouvel éclairage aux règles de l'équilibre mécanique ou du quasi-équilibre sous mouvements lents, principe des travaux virtuels et loi des actions mutuelles;
- aux actions dynamiques (*en utilisant le produit tensoriel symétrique des forces d'accélération par les vitesses des points matériels auxquels elles s'appliquent*), ce qui pourrait donner un nouvel éclairage à la dynamique de ce type de matériaux.

## REMERCIEMENTS

# ANNEXES

## ANNEXES A LA SECTION 1

### *Annexe 1-1 Element de volume pseudo-continu equivalent*

Reprenant la Figure 1.2, désignons par $\Delta x$, $\Delta y$, et $\Delta z$, les dimensions du volume parallélipipédique sur les axes 1,2, et 3. Choisissant une représentation simple et coaxiale, il vient:

-par l'équilibre des forces
$$\begin{cases} \mathbf{s_1} \Delta y \Delta z = f_1 = \|\vec{f}\| \cos(\tfrac{p}{4} - \tfrac{y}{2}) \\ \mathbf{s_2} \Delta x \Delta z \ldots\ldots\ldots\ldots \text{indetermine} \\ \mathbf{s_3} \Delta x \Delta y = f_3 = -\|\vec{f}\| \sin(\tfrac{p}{4} - \tfrac{y}{2}) \end{cases}$$

-par l'équivalence des mouvements
$$\begin{cases} \dot{\mathbf{e}}_1 \Delta x = v_1 = \|\vec{v}\| \cos(\tfrac{p}{4} - \tfrac{y}{2}) \\ \dot{\mathbf{e}}_2 \Delta y = 0 \\ \dot{\mathbf{e}}_3 \Delta z = -v_3 = -\|\vec{v}\| \sin(\tfrac{p}{4} - \tfrac{y}{2}) \end{cases}$$

Notant que le volume du parallélipipède est le produit $\Delta x \cdot \Delta y \cdot \Delta z$, on obtient les résultats de la section 1-2-2.

### *Annexe 1-2 Propriétés essentielles de la norme tensorielle $N_I$*

- La fonction $N_I(\underline{\mathbf{p}}) = |\mathbf{p}_1| + |\mathbf{p}_2| + |\mathbf{p}_3|$ est *définie, positive*, vérifie $N_I(\mathbf{I} \cdot \underline{\mathbf{p}}) = |\mathbf{I}| N_I(\underline{\mathbf{p}})$, et la *boule unité*, dans un repère donné, est un octaèdre régulier, dont les sommets sont sur les axes, aux abscisses $\pm 1$.

- Montrons que, pour tout ensemble de tenseurs $\underline{\mathbf{p}}$, du type défini en section 1-2 de l'article, soient $\underline{\mathbf{p}}(1)..\underline{\mathbf{p}}(k)...\underline{\mathbf{p}}(n)$, d'orientations mutuelles *quelconques*, alors :

    a) **on a toujours** $\quad N_I \left\{ \sum_{1 \leq k \leq n} \underline{\mathbf{p}}(k) \right\} \leq \sum_{1 \leq k \leq n} N_{\|}\left\{ \underline{\mathbf{p}}(k) \right\}$
    
    b) **les seules solutions réalisant l'*égalité* sont :**
    -soit toutes les directions propres portant $\mathbf{p}^+(1),..\mathbf{p}^+(k),..\mathbf{p}^+(n)$ **sont confondues ;**
    -soit toutes les directions propres portant $\mathbf{p}^-(1),.. \mathbf{p}^-(k),...\mathbf{p}^-(n)$ **, sont confondues ;**
    -soit les deux conditions précédentes sont réalisées simultanément.

*Avant d'aller plus avant, remarquons que la démonstration faite ici sur des sommations discrètes représentant une population finie de tenseurs élémentaires $\underline{\mathbf{p}}(m/n)$, s'étendrait aussi bien à l'intégration sur une population représentée par une distribution statistique continue.*

- Notons que la somme $\underline{\mathbf{P}} = \sum_{1 \leq k \leq n} \underline{\mathbf{p}}(k)$ est un tenseur à trace positive (somme de tenseurs à trace positive) et possède donc au moins une valeur propre positive. Ordonnons les valeurs propres de $\underline{\mathbf{P}}$ par valeurs décroissantes, le tableau suivant résume toutes les combinaisons possibles, ainsi que leur correspondance avec les « cas » analysés ci-dessous :

| Signe de $P_1$ | Signe $P_2$ | Signe $P_3$ | Cas d'analyse |
|---|---|---|---|
| + | - | - | A |
| + | 0 | - | A **et** B |
| + | + | - | B |
| + | + | 0 | C |
| + | + | + | C |

## A] $\underline{P}$ a une seule valeur propre positive $P_1 = P^+$

- Appelons $\vec{n}^+$ la direction portant $P^+$

  Dans les repères propres de <u>chacun</u> des $\underline{p}(k)$, prenons comme convention de numérotation $\mathbf{p}_1(k) = \mathbf{p}^+(k) > 0$, $\mathbf{p}_2(k) = 0$, $\mathbf{p}_3(k) = \mathbf{p}^-(k) < 0$,

  et notons les coordonnées de la direction $\vec{n}^+$ : $\begin{cases} \sin a_k \cos b_k \\ \cos a_k \\ \sin a_k \sin b_k \end{cases}$

- En sommant les contributions de tous les $\underline{p}(k)$ suivant $\vec{n}^+$, il vient:

$$P_1 = P^+ = \sum_{1 \leq k \leq n} \left\{ \mathbf{p}^+(k) \sin^2 a_k \cos^2 b_k + \mathbf{p}^-(k) \sin^2 a_k \sin^2 b_k \right\} \quad .....(a.1)$$

dans l'expression (a.1), les seuls termes <u>positifs</u> sont ceux correspondant aux $\mathbf{p}^+(k)$, tous les autres sont négatifs; de plus, les coefficients trigonométriques sont $\leq 1$, d'où :

$$0 \leq P^+ \leq \sum_{1 \leq k \leq n} \mathbf{p}^+(k) \quad ..............................(a.2).$$

- Par définition $P^+ = \tfrac{1}{2}\left[Tr\{\underline{P}\} + N_{|}\{\underline{P}\}\right]$, de même pour les $\mathbf{p}^+(k)$.

  Comme $\underline{P} = \sum_{1 \leq k \leq n} \underline{p}(k)$, on a : $Tr\{\underline{P}\} = \sum_{1 \leq k \leq n} Tr\{\underline{p}(k)\}$

  Remplaçant terme à terme dans (a.2), et tenant compte de la relation sur les traces, vient l'inégalité :

$$N_{|}\left\{\sum_{1 \leq k \leq n} \underline{p}(k)\right\} \leq \sum_{1 \leq k \leq n} N_{\|}\{\underline{p}(k)\} \quad .........(a.3)$$

- Une solution assurant l' *égalité dans (a.3)*, nécessiterait dans (a.1) la réalisation simultanée des conditions :

  -les coefficients trigonométriques des $\mathbf{p}^-(k)$ sont tous nuls ;

  -les coefficients trigonométriques des $\mathbf{p}^+(k)$ sont tous égaux à **1**.

  La seule et unique solution est que pour tous les $k$ on ait $\sin a_k \cos b_k = \pm 1$, c'est à dire que $\vec{n}^+$ et la direction portant $\mathbf{p}^+(k)$ coïncident.

  Donc, *toutes les directions propres portant* $\mathbf{p}^+(1)$, .. $\mathbf{p}^+(k)$ ... $\mathbf{p}^+(n)$, *sont confondues avec celle portant* $P^+$.

## B] $\underline{\mathbf{P}}$ a une seule valeur propre négative $\mathbf{P}_3 = \mathbf{P}^-$

- Par un raisonnement symétrique du précédent, mais appuyé ici sur les valeurs propres ***négatives***, on trouve ici :
$$\sum_{1 \leq k \leq n} \mathbf{p}^-(k) \leq \mathbf{P}^- \leq 0$$

D'où l'inégalité cherchée.

Analysant les conditions pouvant assurer l'**égalité** dans (a.3), on trouve ici que *toutes les directions propres portant* $\mathbf{p}^-(1), .. \mathbf{p}^-(k)... \mathbf{p}^-(n)$, *sont confondues avec celle portant* $\mathbf{P}^-$.

## C] Toutes les valeurs propres de $\underline{\mathbf{P}}$ sont positives ou nulles.

- Dans ce cas on vérifie $Tr\{\underline{\mathbf{P}}\} = N_{\|}\{\underline{\mathbf{P}}\}$, par ailleurs on a toujours
$$Tr\{\underline{\mathbf{P}}\} = \sum_{1 \leq k \leq n} Tr\{\underline{\mathbf{p}}(k)\}.$$

Chacun des $\underline{\mathbf{p}}(k)$ ayant des valeurs propres de signe différents,
$$Tr\{\underline{\mathbf{p}}(k)\} < N_{\|}\{\underline{\mathbf{p}}(k)\}$$

Rapprochant ces relations, on obtient :
$$N_{\|}\left\{\sum_{1 \leq k \leq n} \underline{\mathbf{p}}(k)\right\} < \sum_{1 \leq k \leq n} N_{\|}\{\underline{\mathbf{p}}(k)\}$$

- Dans ce cas d'inégalité ***stricte***, il n'y a ***aucune*** solution qui pourrait assurer l'égalité.

## ANNEXES A LA SECTION 2

### *2-1 PROPRIETES DE LA FONCTION* $\mathbf{\mathit{E}}(A) = R(A) \sum_A N_{\|}\{\underline{\mathbf{p}}(m/n)\}$

- A un coefficient positif près, la fonction $\mathbf{\mathit{E}}(A)$ a les propriétés de

$f(A) = \sum_A N_{\|}\{\underline{\mathbf{p}}(m/n)\} - N_{\|}\left\{\sum_A \underline{\mathbf{p}}(m/n)\right\}$ fonction toujours positive (propriétés de la norme), et qui s'annule si A est en mode de dissipation minimale (*R*=0).

- Si A est formé par réunion de deux sous-amas $A_1$ et $A_2$, on a:
$$f(A_1 \cup A_2) = \sum_{A_1 \cup A_2} N_{\|}\{\underline{\mathbf{p}}(m/n)\} - N_{\|}\left\{\sum_{A_1 \cup A_2} \underline{\mathbf{p}}(m/n)\right\}$$

-le premier terme de cette relation est toujours additif
$$\sum_{A_1 \cup A_2} N_{\|}\{\underline{\mathbf{p}}(m/n)\} = \sum_{A_1} N_{\|}\{\underline{\mathbf{p}}(m/n)\} + \sum_{A_2} N_{\|}\{\underline{\mathbf{p}}(m/n)\}$$

- par contre, pour le deuxième terme, d'après le résultat de l'Annexe 1-2:
$$N_{\|}\left\{\sum_{A_1 \cup A_2} \underline{\mathbf{p}}(m/n)\right\} \leq N_{\|}\left\{\sum_{A_1} \underline{\mathbf{p}}(m/n)\right\} + N_{\|}\left\{\sum_{A_2} \underline{\mathbf{p}}(m/n)\right\}$$

Enfin, considérant les tenseurs de puissance $\underline{\mathbf{P}}(A_1)$ et $\underline{\mathbf{P}}(A_2)$ attachés aux sous-amas granulaires $A_1$ et $A_2$, la condition qui, dans l'inégalité ci-dessus, réaliserait l'égalité, est que:
$$N_{\|}\{\underline{\mathbf{P}}(A_1 \cup A_2)\} = N_{\|}\{\underline{\mathbf{P}}(A_1) + \underline{\mathbf{P}}(A_2)\} = N_{\|}\{\underline{\mathbf{P}}(A_1)\} + N_{\|}\{\underline{\mathbf{P}}(A_2)\}$$

D'après le résultat de l'Annexe 1-2, ceci ne sera possible que si $\underline{\mathbf{P}}(A_1)$ et $\underline{\mathbf{P}}(A_2)$ ont le même nombre de valeurs propres positives, et si les directions principales correspondantes coïncident. Ceci exclut toute réalimentation entre les sous-amas $A_1$ et $A_2$.

- On aura donc toujours $f(A_1 \cup A_2) \geq f(A_1) + f(A_2)$, l'égalité ne pouvant être assurée qu'en l'absence de réalimentation entre $A_1$ et $A_2$.
- Enfin, si les sous-amas granulaires sont chacun en dissipation minimale, et sans réalimentation entre eux, les termes de l'inégalité précédente deviennent tous nuls.

## 2-2- DISTRIBUTIONS DES ORIENTATIONS DES TENSEURS ELEMENTAIRES $\underline{\mathbf{p}}(m/n)$, VERIFIANT LA DISSIPATION MAXIMALE.

Dans ce qui suit, nous analyserons d'abord le cas monoaxial, puis la famille remarquable de solutions particulières composées par combinaisons de mouvements des Modes I et II de la dissipation minimale, de même axe mais contraposés, en nous plaçant dans le repère propre du tenseur résultant $\underline{\mathbf{P}}$.

### A] Distributions monoaxiales $\mathbf{P}_1 = \mathbf{P}^+ > 0$, $\mathbf{P}_2 = 0$, $\mathbf{P}_3 = 0$

a) <u>solutions vérifiant la propriété sur chacun des $\underline{\mathbf{p}}(m/n)$</u> :

dans le repère propre du tenseur $\underline{\mathbf{P}}$, on a donc $\begin{cases} p_{11}(m/n) > 0 \\ p_{22}(m/n) = 0 \\ p_{33}(m/n) = 0 \end{cases}$

Pour chacun des $\underline{\mathbf{p}}(m/n)$, le plan 2-3 est donc formé par la normale au plan de contact et l'axe neutre, ou encore est le symétrique de celui-ci par rapport à la direction portant $\underline{\mathbf{p}}^+(m/n)$ (voir section 1-2-3, sur les directions de puissance purement tangentielle). Les directions des vitesses relatives de glissement sont donc: soit <u>parallèles à l'axe 1</u>, soit <u>inclinées à $\dfrac{\boldsymbol{p}}{2} - \boldsymbol{y}$ par rapport à cet axe (figure 2-8)</u>.

b) <u>solutions vérifiant la propriété seulement sur l'ensemble des $\underline{\mathbf{p}}(m/n)$</u> :

   b1) *distributions coaxiales*

Considérons une distribution $D'$ correspondant à un mouvement de dissipation minimale de Mode I, d'axe 1, dont les tenseurs élémentaires $\underline{\mathbf{p}}'(m/n)$ vérifient donc $\begin{cases} p'_{11} = \mathbf{p}'^+ \\ p'_{22} + p'_{33} = \mathbf{p}'^- \end{cases}$.

Sommons-la avec une distribution $D''$, correspondant à un mouvement de dissipation minimale en Mode II, toujours d'axe 1, obtenue en faisant correspondre à chaque tenseur élémentaire $\underline{\mathbf{p}}'(m/n)$ de $D'$, un tenseur élémentaire $\underline{\mathbf{p}}''(m/n)$, obtenu par rotation et similitude sur $\underline{\mathbf{p}}'(m/n)$, tel que

$$\begin{cases} p''_{11} = -\left(\dfrac{1-\sin\boldsymbol{y}}{1+\sin\boldsymbol{y}}\right)^2 \cdot p'_{11} = \left(\dfrac{1-\sin\boldsymbol{y}}{1+\sin\boldsymbol{y}}\right) \cdot \mathbf{p}'^- = \mathbf{p}''^- \\ \quad p''_{22} = -p'_{22} \quad \text{idem pour} \quad p''_{33} \\ \text{d'où} \quad p''_{22} + p''_{33} = -\mathbf{p}'^- = \left(\dfrac{1-\sin\boldsymbol{y}}{1+\sin\boldsymbol{y}}\right) \cdot \mathbf{p}'^+ = \mathbf{p}''^+ \end{cases}$$

Dans la distribution résultante $D = D' \bigcup D''$ les sommes $\underline{p}'(m/n) + \underline{p}''(m/n)$ vérifient:

$$\begin{cases} p'_{11} + p''_{11} = \mathbf{p}'^+ + \mathbf{p}''^- = \mathbf{p}'^+ + \left(\dfrac{1-\sin\boldsymbol{\varphi}}{1+\sin\boldsymbol{\varphi}}\right)\mathbf{p}'^- > 0 \\ p'_{22} + p''_{22} = 0 \quad \text{et} \quad p'_{33} + p''_{33} = 0 \end{cases}$$

Le tenseur résultant $\underline{\mathbf{P}}$ correspondant à $D$ vérifie donc bien les propriétés $\mathbf{P}_1 = \mathbf{P}^+ > 0$, $\mathbf{P}_2 = 0$, $\mathbf{P}_3 = 0$

Notons qu'il n'est pas nécessaire que les distributions $D'$ et $D''$ soient chacune symétriques, il suffit que la similitude entre les deux soit respectée.

Remarquons aussi que la décomposition précédente du tenseur résultant $\underline{\mathbf{P}}$, peut s'écrire symboliquement: $\underline{\mathbf{P}} = \begin{pmatrix}+\\0\\0\end{pmatrix} = \begin{pmatrix}+\\-\\-\end{pmatrix} + \begin{pmatrix}-\\+\\+\end{pmatrix}$ , ce qui suggère d'autres combinaisons, telles que

$\underline{\mathbf{P}} = \begin{pmatrix}+\\0\\0\end{pmatrix} = \begin{pmatrix}-\\+\\-\end{pmatrix} + \begin{pmatrix}+\\-\\+\end{pmatrix}$ correspondant à une décomposition analogue à la précédente, mais axées sur la direction propre n°2, orthogonale à l'axe de notre tenseur résultant uniaxial $\underline{\mathbf{P}}$, et dont les proportions seront différentes. Nous en verrons une autre en b3) ci-dessous, correspondant à

$$\underline{\mathbf{P}} = \begin{pmatrix}+\\0\\0\end{pmatrix} = \begin{pmatrix}+\\+\\-\end{pmatrix} + \begin{pmatrix}+\\-\\+\end{pmatrix}$$

b2) *distributions en déformation plane correspondantes*

Les distributions précédentes ( a),et b1)) ne sont pas nécéssairement à symétrie de révolution autour de l'axe 1, il suffit qu'elles soient formées de paires de tenseurs élémentaires qui se compensent deux à deux ; on peut particulariser la direction de l'axe neutre sur l'axe 2, obtenant ainsi des combinaisons de distributions en déformation plane, vérifiant la condition monoaxiale.

b3) *distributions d'axes orthogonaux*

Considérons une distribution $D'$ correspondant à un mouvement de dissipation minimale de Mode II, d'axe 3, dont le tenseur de puissance ***résultant*** est $\underline{\mathbf{P}}' = \begin{cases} P'_1 = (1-a)\mathbf{P}'^+ \\ P'_2 = a\mathbf{P}'^+ \\ P'_3 = -\dfrac{(1-\sin\boldsymbol{\varphi})}{(1+\sin\boldsymbol{\varphi})}\mathbf{P}'^+ \end{cases}$ avec $\left(\dfrac{1-\sin\boldsymbol{\varphi}}{1+\sin\boldsymbol{\varphi}}\right)^2 \leq a \leq 1$

Sommons –la avec une distribution $D''$, correspondant à un mouvement de dissipation minimale en Mode II, d'axe 2, dont le tenseur de puissance *résultant* est $\underline{\mathbf{P}}'' = \begin{cases} P_1'' = \left[ a\left(\dfrac{1+\sin \boldsymbol{y}}{1-\sin \boldsymbol{y}}\right) - \left(\dfrac{1-\sin \boldsymbol{y}}{1+\sin \boldsymbol{y}}\right) \right] \mathbf{P}'^+ \\ P_2'' = -a\mathbf{P}'^+ \\ P_3'' = \dfrac{(1-\sin \boldsymbol{y})}{(1+\sin \boldsymbol{y})} \mathbf{P}'^+ \end{cases}$

Dans la distribution résultante $D = D' \cup D''$ le tenseur résultant $\underline{\mathbf{P}} = \underline{\mathbf{P}}' + \underline{\mathbf{P}}''$ correspondant à D vérifie donc bien les propriétés $\mathbf{P_1} = \mathbf{P}^+ > 0$, $\mathbf{P_2} = 0$, $\mathbf{P_3} = 0$.

Donc une distribution monoaxiale de dissipation Maximale peut être obtenue comme résultant de deux composantes en dissipation minimale, toutes deux de Mode II, et d'axes orthogonaux.

  b4) *Distributions quelconques*

Il faut revenir aux relations de base exprimant que deux des valeurs propres du tenseur résultant $\underline{\mathbf{P}}$, sont nulles (relation (a.1) de l'annexe 1-2), ce qui laisse un grand nombre de degrés de liberté à ces distributions quelconques.

**B] Solutions particulières combinaisons de mouvements coaxiaux des modes de dissipation minimale.**

Considérons donc la composition de deux solutions en dissipation minimale, Modes I et II, de même axe 1, dans le repère propre du tenseur résultant $\underline{\mathbf{P}}$, ces deux distributions ne sont pas nécessairement à symétrie de révolution autour de l'axe 1.

Dans la puissance dissipée totale $D_{Tot}$, les participations $D_I$ et $D_{II}$ de chacune des composantes peuvent etre notées : $\begin{cases} D_I = \dfrac{(1-\boldsymbol{I})}{2} D_{Tot} \\ D_{II} = \dfrac{(1+\boldsymbol{I})}{2} D_{Tot} \end{cases}$

Les puissances $\mathbf{P}^+$ et $\mathbf{P}^-$ de chacune des composantes sont alors,

pour la composante de Mode I : $\begin{cases} \mathbf{P}_I^+ = \dfrac{(1+\sin \boldsymbol{y})}{2\sin \boldsymbol{y}} \dfrac{(1-\boldsymbol{I})}{2} D_{Tot} \quad \text{sur l'axe 1} \\ \mathbf{P}_I^- = -\dfrac{(1-\sin \boldsymbol{y})}{2\sin \boldsymbol{y}} \dfrac{(1-\boldsymbol{I})}{2} D_{Tot} \end{cases}$ et pour la composante de Mode II: $\begin{cases} \mathbf{P}_{II}^+ = \dfrac{(1+\sin \boldsymbol{y})}{2\sin \boldsymbol{y}} \dfrac{(1+\boldsymbol{I})}{2} D_{Tot} \\ \mathbf{P}_{II}^- = -\dfrac{(1-\sin \boldsymbol{y})}{2\sin \boldsymbol{y}} \dfrac{(1+\boldsymbol{I})}{2} D_{Tot} \quad \text{sur l'axe 1} \end{cases}$

Sur l'axe 1 vont s'ajouter les puissances $\mathbf{P}_I^+$ et $\mathbf{P}_{II}^-$, tandis que sur l'ensemble des deux autres axes, vont s'ajouter les puissances $\mathbf{P}_{II}^+$ et $\mathbf{P}_I^-$, qui se trouvent réparties sur ces deux axes.

 Les valeurs propres du tenseur résultant vérifient donc $\begin{cases} \mathbf{P_1} = \mathbf{P}_I^+ + \mathbf{P}_{II}^- = \tfrac{1}{2}\left(1 - \dfrac{\boldsymbol{I}}{\sin \boldsymbol{y}}\right) D_{Tot} \\ \mathbf{P_2} + \mathbf{P_3} = \mathbf{P}_{II}^+ + \mathbf{P}_I^- = \tfrac{1}{2}\left(1 + \dfrac{\boldsymbol{I}}{\sin \boldsymbol{y}}\right) D_{Tot} \end{cases}$

Ces relations font apparaître quatre cas.

a) $\boldsymbol{I} = -\sin \boldsymbol{y}$ Dans ce cas les valeurs propres $\mathbf{P_2}$ et $\mathbf{P_3}$ sont neutralisées dans le tenseur résultant (car leur somme est nulle, alors qu'elles sont toutes deux positives ou nulles, dans ce cas de la dissipation maximale). Il s'agit donc de la situation **monoaxiale** (seul l'axe 1 est actif), et les participations des deux composantes de Modes I et II dans la dissipation totale sont:
$$\begin{cases} D_I = \frac{(1+\sin\mathbf{y})}{2} D_{Tot} \\ D_{II} = \frac{(1-\sin\mathbf{y})}{2} D_{Tot} \end{cases}$$

b) $\mathbf{l} = +\sin\mathbf{y}$ Dans ce cas la valeur propre $\mathbf{P_1}$ est neutralisée dans le tenseur résultant. Il s'agit donc de la situation **biaxiale** (seuls les axes 2 et 3 sont actifs), et les participations des deux composantes de Modes I et II dans la dissipation totale sont:
$$\begin{cases} D_I = \frac{(1-\sin\mathbf{y})}{2} D_{Tot} \\ D_{II} = \frac{(1+\sin\mathbf{y})}{2} D_{Tot} \end{cases}$$

c) $-\sin\mathbf{y} < \mathbf{l} < \sin\mathbf{y}$ Dans ce cas toutes les valeurs propres du tenseur résultant peuvent être positives. Il s'agit donc de situations **tridimensionnelles** (3 axes actifs), l' isotropie a lieu, lorsque les distributions sont à symétrie axiale d'axe 1, pour $\mathbf{l} = \frac{1}{3}\sin\mathbf{y}$.

Pour ce cas d'isotropie, les participations des composantes de modes sont:
$$\begin{cases} D_I = \frac{(3-\sin\mathbf{y})}{6} D_{Tot} \\ D_{II} = \frac{(3+\sin\mathbf{y})}{6} D_{Tot} \end{cases}$$

d) $-1 < \mathbf{l} < -\sin\mathbf{y}$ ou $\sin\mathbf{y} < \mathbf{l} < 1$ On sort ici des solutions de dissipation Maximale pour entrer dans les solutions intermédiaires (0<R<1), qui sont, dans le premier cas, tridimensionnelles à dominante Mode I, et dans le deuxième cas, tridimensionnelles à dominante Mode II.

Enfin, il est utile de résumer les éléments précédents dans le tableau suivant, qui donne les valeurs des participations de composantes de modes, pour des valeurs type de l'angle de friction $\mathbf{y}$.

### DISSIPATION MAXIMALE

Solutions particulières par compositions de mouvements de Modes de dissipation minimale, de même axe mais contraposés.
*Participations de chacun des Modes dans la dissipation totale*

|  | Solutions monoaxiales | | Solutions biaxiales | | Solutions tridimensionnelles Cas isotrope | |
|---|---|---|---|---|---|---|
| **Friction $\mathbf{y}$ (°)** | Mouvement principal: Mode I $\left(\frac{1+\sin\mathbf{y}}{2}\right)$ | Mouvement secondaire: Mode II $\left(\frac{1-\sin\mathbf{y}}{2}\right)$ | Mouvement principal: Mode II $\left(\frac{1+\sin\mathbf{y}}{2}\right)$ | Mouvement secondaire: Mode I $\left(\frac{1-\sin\mathbf{y}}{2}\right)$ | Mouvement principal: Mode II $\left(\frac{3+\sin\mathbf{y}}{6}\right)$ | Mouvement secondaire: Mode I $\left(\frac{3-\sin\mathbf{y}}{6}\right)$ |
| 20 | 67% | 33% | 67% | 33% | 56% | 44% |
| 30 | 75% | 25% | 75% | 25% | 58% | 42% |
| 40 | 82% | 18% | 82% | 18% | 61% | 39% |
| 50 | 88% | 12% | 88% | 12% | 63% | 36% |

**ANNEXE A LA SECTION 3**

*ANNEXE 3-1: RELATIONS CONTRAINTES-DILATANCES DE ROWE*

- Pour retrouver les relations de Rowe à partir de l'équation de dissipation du milieu continu équivalent (3.1), il est utile d'introduire dans celle-ci les variables auxiliaires $\boldsymbol{a}_i = \dfrac{\dot{\varepsilon}_i}{|\dot{\varepsilon}_i|}$, *pour* $\dot{\varepsilon}_i \neq 0$

  L'équation de dissipation (3.1) devient alors $\sum_i (1 - \boldsymbol{a}_i \sin \boldsymbol{y}) \sigma_i \dot{\varepsilon}_i = 0$

- Sous **contraintes axissymétriques**, donnant l'indice 1 aux grandeurs axiales, $\sigma_2 = \sigma_3$, et dans l'hypothèse où $\dot{\varepsilon}_2$ et $\dot{\varepsilon}_3$ ont le même signe (alors $\boldsymbol{a}_2 = \boldsymbol{a}_3$), l'équation précédente s'écrit: $(1 - \boldsymbol{a}_1 \sin \boldsymbol{y}) \sigma_1 \dot{\varepsilon}_1 + (1 - \boldsymbol{a}_3 \sin \boldsymbol{y}) \sigma_3 (\dot{\varepsilon}_2 + \dot{\varepsilon}_3) = 0$

  Remarquons que $-(\dot{\varepsilon}_2 + \dot{\varepsilon}_3)/\dot{\varepsilon}_1 = 1 - \dot{\varepsilon}_v / \dot{\varepsilon}_1$, l'équation ci-dessus devient:

  $$\frac{\sigma_1}{\sigma_3} = \left(1 - \frac{\dot{\varepsilon}_v}{\dot{\varepsilon}_1}\right) \cdot \frac{(1 - \boldsymbol{a}_3 \sin \boldsymbol{y})}{(1 - \boldsymbol{a}_1 \sin \boldsymbol{y})}$$

- Suivant que la sollicitation est en mode de type I ($\boldsymbol{a}_1 = +1$) ou de type II ($\boldsymbol{a}_1 = -1$), comme en général $\boldsymbol{a}_1 \boldsymbol{a}_3 = -1$ (nous excluons ici le cas de la dissipation Maximale), on obtient directement les _**relations de Rowe**_ :

  - écrasement axial (mode type I)   $\dfrac{\sigma_1}{\sigma_3} = \left(1 - \dfrac{\dot{\varepsilon}_v}{\dot{\varepsilon}_1}\right) \cdot \dfrac{(1 + \sin \boldsymbol{y})}{(1 - \sin \boldsymbol{y})}$

  - écrasement radial (mode type II)   $\dfrac{\sigma_1}{\sigma_3} = \left(1 - \dfrac{\dot{\varepsilon}_v}{\dot{\varepsilon}_1}\right) \cdot \dfrac{(1 - \sin \boldsymbol{y})}{(1 + \sin \boldsymbol{y})}$

- Dans le cas de la **déformation plane**, il suffit de poser $\dot{\varepsilon}_2 = 0$ dans les formules précédentes, et les deux équations finales se trouvent conservées.

Remarquons que dans les relations ci-dessus, plus la dilatance est forte (c'est à dire plus $\dot{\varepsilon}_v$ est <0), plus fort est l'écart entre les contraintes principales.

*ANNEXE 3-2: CRITÈRE DE COULOMB À L'ÉTAT CRITIQUE*

Lorsque le volume spécifique devient stationnaire ($\dot{\varepsilon}_v = 0$), alors les relations ci-dessus deviennent:

$$\frac{Sup(\sigma_1, \sigma_2, \sigma_3)}{Inf(\sigma_1, \sigma_2, \sigma_3)} = \frac{(1 + \sin \boldsymbol{y})}{(1 - \sin \boldsymbol{y})} = \tan^2\left(\frac{\boldsymbol{p}}{4} + \frac{\boldsymbol{y}}{2}\right)$$

Cette relation est celle du Critère de Coulomb, obtenue ici vis à vis de situations particulières (symétrie axiale en contraintes ou déformation plane).
Vis à vis de situations tridimensionnelles en général, on peut montrer qu'à l'état critique:
  - le Critère de Coulomb est bien la solution de (3.1) qui donne à la fois le minimum de résistance (en terme de rapport des contraintes principales extrêmes), et le minimum minimorum de la dissipation d'énergie, pour les conditions aux limites usuelles,
  - qu'il est associé à de la déformation plane sur la direction de la contrainte principale intermédiaire.

*ANNEXE 3-3: DOMAINE CARACTÉRISTIQUE*

Si l'état de contraintes demeure à l'intérieur de la surface du Critère de Coulomb pendant le mouvement, alors les relations ci-dessus impliquent que les déformations en volume sont toujours en contraction ($\dot{\varepsilon}_v > 0$), ce qui correspond au concept du "domaine caractéristique".

Vis à vis de situations tridimensionnelles en général, on peut montrer que cette propriété demeure vraie.

Remarquons enfin que lorsque l'état de contraintes est voisin de l'isotropie (déviateur "petit" devant la pression isotrope), alors l'équation de dissipation se simplifie en une équation de densification: $\dfrac{\dot{\varepsilon}_v}{\sum |\dot{\varepsilon}_i|} = \sin y$, qui indique que pour toute déformation ($\sum |\dot{\varepsilon}_i| \neq 0$) alors le volume spécifique diminue ($\dot{\varepsilon}_v > 0$).

## *ANNEXE 3-4: PASSAGE À LA MÉCANIQUE DES FLUIDES*

Lorsque $y$ tend vers zéro, le Critère de Coulomb indique que le milieu peut se déformer indéfiniment à volume stationnaire sous contraintes qui tendent vers l'isotropie: la résistance au cisaillement disparaît.

Réciproquement, sous état de contraintes voisin de l'isotropie, l'équation de densification vue ci-dessus, indique que le volume spécifique devient stationnaire lorsque $y$ tend vers zéro.

Enfin, le deuxième membre de l'équation de dissipation (3.1) tend aussi vers zéro, un raisonnement classique (voir [18]) déduit alors que si le volume spécifique est constant au cours de la déformation, à dissipation nulle, alors les contraintes sont nécessairement isotropes.

On obtient bien le comportement d'un "fluide parfait incompressible"

# Références